\documentclass[12pt]{article}

\usepackage{algorithmic}
\usepackage{algorithm}
\usepackage{amsmath}
\usepackage{times}
\usepackage{graphicx, subfigure}
\usepackage{epsfig}
\usepackage{epstopdf}
\usepackage{color}
\usepackage{multirow}

\usepackage[authoryear]{natbib}
\usepackage{rotating}
\usepackage{bbm}
\usepackage{latexsym}
\DeclareGraphicsExtensions{-eps-converted-to.pdf,.png}

\usepackage{enumerate}
\usepackage{setspace}
\usepackage{graphicx,psfrag,epsfig,multirow}
\usepackage{amsmath,amsthm,amssymb}

\renewcommand{\epsilon}{\varepsilon}
\renewcommand{\Sigma}{\varSigma}

\usepackage{psfig}
\usepackage{epsfig}
\usepackage{epstopdf}
\usepackage[leftcaption]{sidecap}

\usepackage{hyperref}

\usepackage{url}

\usepackage{algorithmic}
\usepackage{algorithm}

\usepackage{tikz}
\usetikzlibrary{shapes.symbols}
\usetikzlibrary{decorations.pathmorphing}
\usetikzlibrary{shapes,shadows}
\usepackage{graphics}
\usepackage{pdfsync}
\usepackage{sidecap}

\usepackage{caption}

\newcommand{\captionfonts}{\normalsize}

\makeatletter
\long\def\@makecaption#1#2{%
  \vskip\abovecaptionskip
  \sbox\@tempboxa{{\captionfonts #1: #2}}%
  \ifdim \wd\@tempboxa >\hsize
    {\captionfonts #1: #2\par}
  \else
    \hbox to\hsize{\hfil\box\@tempboxa\hfil}%
  \fi
  \vskip\belowcaptionskip}
\makeatother
\begin{document}
\hspace{13.9cm}1

\ \vspace{20mm}\\

\begin{center}
{\LARGE A Semiparametric Bayesian Model for Detecting Synchrony Among Multiple Neurons}
\end{center}

\ \\
{\bf \large Babak Shahbaba$^{\displaystyle 1}$, Bo Zhou$^{\displaystyle 1}$, Shiwei Lan$^{\displaystyle 1}$, Hernando Ombao$^{\displaystyle 1}$, David Moorman$^{\displaystyle 2}$, Sam Behseta$^{\displaystyle 3}$}\\
{$^{\displaystyle 1}$Department of Statistics, UC Irvine, CA.}\\
{$^{\displaystyle 2}$Department of Psychology, University of Massachusetts Amherst, MA.}\\
{$^{\displaystyle 3}$Department of Mathematics, California State University, Fullerton, CA.}\\
%

%\ \\[-2mm]

\thispagestyle{empty}
\markboth{}{NC instructions}
\ \vspace{-0mm}\\
%
%Abstract
\begin{center} {\bf Abstract} \end{center}
We propose a scalable semiparametric Bayesian model to capture dependencies among multiple neurons by detecting their co-firing (possibly with some lag time) patterns over time. After discretizing time so there is at most one spike at each interval, the resulting sequence of 1's (spike) and 0's (silence) for each neuron is modeled using the logistic function of a continuous latent variable with a Gaussian process prior. For multiple neurons, the corresponding marginal distributions are coupled to their joint probability distribution using a parametric copula model. The advantages of our approach are as follows: the nonparametric component (i.e., the Gaussian process model) provides a flexible framework for modeling the underlying firing rates; the parametric component (i.e., the copula model) allows us to make inference regarding both contemporaneous and lagged relationships among neurons; using the copula model, we construct multivariate probabilistic models by separating the modeling of univariate marginal distributions from the modeling of dependence structure among variables; our method is easy to implement using a computationally efficient sampling algorithm that can be easily extended to high dimensional problems. Using simulated data, we show that our approach could correctly capture temporal dependencies in firing rates and identify synchronous neurons. We also apply our model to spike train data obtained from prefrontal cortical areas in rat's brain.

%%%%%%%%%%%

\section{Introduction}\label{introduction}

Neurophysiological studies commonly involve modeling a sequence of spikes (action potentials) over time, known as a \emph{spike train}, for each neuron. However, complex behaviors are driven by networks of neurons instead of a single neuron. In this paper, we propose a flexible, yet robust semiparametric Bayesian method for capturing temporal cross-dependencies among multiple neurons by simultaneous modeling of their spike trains. In contrast to most existing methods, our approach provides a flexible, yet powerful and scalable framework that can be easily extended to high dimensional problems.

For many years preceding ensemble recording, neurons were recorded successively and then combined into synthetic populations based on shared timing. Although this technique continues to produce valuable information to this day \citep{meyer11}, investigators are gravitating more and more towards simultaneous recording of multiple single neurons \citep{miller08}. A major reason that multiple-electrode recording techniques have been embraced is because of the ability to identify the activity and dynamics of populations of neurons simultaneously.  It is widely appreciated that groups of neurons encode variables and drive behaviors \citep{buzsaki10}. 

Early analysis of simultaneously recorded neurons focused on correlation of activity across pairs of neurons using cross correlation analyses \citep{narayanan09} and analyses of changes in correlation over time, i.e., 
by using a joint peristimulus time histogram (JPSTH) \citep{gerstein69} or rate correlations \citep{narayanan09}.  
Similar analyses can be performed in the frequency domain by using coherence analysis of neuron pairs using 
Fourier-transformed neural activity \citep{brown04}. These methods attempt to distinguish \emph{exact synchrony} or \emph{lagged synchrony} between a pair of neurons. Subsequently, a class of associated methods were developed for addressing the question of whether exact or lagged synchrony in a pair of neurons is merely due to chance. Later, to test the statistical significance of synchrony, a variety of methods, such as bootstrap confidence intervals, were introduced \citep{harrison13}. 

To detect the presence of conspicuous spike coincidences in multiple neurons, \cite{grun02} proposed a novel method, where such conspicuous coincidences, called \emph{unitary events}, are defined as joint spike constellations that recur more often than what can be explained by chance alone. In their approach, simultaneous spiking events from $N$ neurons are modeled as a joint process composed of $N$ parallel point processes. To test the significance of unitary events, they developed a new method, called joint-surpise, which measures the cumulative probability of finding the same or even larger number of observed coincidences by chance. 

\cite{pillow08} investigate how correlated spiking activity in complete neural populations depends on the pattern of visual simulation. They propose to use a generalized linear model to capture the encoding of stimuli in the spike trains of a neural population. In their approach, a cell's input is presented by a set of linear filters and the summed filter responses are exponantiated to obtain an instantaneous spike rate. The set of filters include a stimulus filter, a post-spike filter (to capture dependencies on history), and a set of coupling filter (to capture dependencies on the recent spiking of other cells). 

Recent developments in detecting synchrony among neurons include models that account for trial to trial variability and the evolving intensity of firing rates between multiple trials. For more discussion on analysis of spike trains, refer to \cite{harrison13, brillinger98, brown04,kass05,west97,rigat06, patnaik08, diekman09, sastry10, kottas12}.

In a recent work, \cite{kelly12} proposed a new method to quantify synchrony. They argue that separating stimulus effects from history effects would allow for a more precise estimation of the instantaneous conditional firing rate.  Specifically, given the firing history $H_{t}$, define $\lambda^A(t|H_t^A)$, $\lambda^{B}(t|H_t^B)$, and $\lambda^{AB}(t|H_t^{AB})$ to be the conditional firing intensities of neuron A, neuron B, and their synchronous spikes respectively. Independence between the two point processes can be examined by testing the null hypothesis $H_0: \zeta(t)=1$, where $\zeta(t)=\frac{\lambda^{AB}(t|H_t^{AB})}{\lambda^A(t|H_t^A)\lambda^B(t|H_t^{B})}$. The quantity $[\zeta(t) - 1]$ can be interpreted as the deviation of co-firing from what is predicted by independence. Note that we still need to model the marginal probability of firing for each neuron. To do this, one could assume that a spike train follows a Poisson process, which is the simplest form of point processes. The main \underline{limitation} of this approach is that it assumes that the number of spikes within a particular time frame follows a Poisson distribution. It is, however, very unlikely that actual spike trains follow this assumption \citep{barbieri01, kass01, reich98, kass05, jacob09}. One possible remedy is to use inhomogeneous Poisson process, which assumes time-varying firing rates. See \cite{brillinger98, brown04,kass05,west97,rigat06, cunningham07, berkes09, kottas10,sacerdote12, kottas12} for more alternative methods for modeling spike trains.

In this paper, we propose a new semiparametric method for neural decoding. We first discretize time so that there is at most one spike within each time interval and let the response variable to be a binary process comprised of 1s and 0s. We then use a continuous latent variable with Gaussian process prior to model the time-varying and history-dependent firing rate for each neuron. The covariance function for the Gaussian process is specified in a way that it creates prior positive autocorrelation for the latent variable so the firing rate could depend on spiking history. For each neuron, the marginal probability of firing within an interval is modeled by the logistic function of its corresponding latent variable. The main advantage of our model is that it connects the joint distribution of spikes for multiple neurons to their marginals by a parametric copula model in order to capture their cross-dependencies. Another advantage is that our model allows for co-firing of neurons after some lag time.

\cite{cunningham07} also assume that the underlying non-negative firing rate is a draw from a Gaussian process. However, unlike the method proposed in this paper, they assume that the observed spike train is a conditionally inhomogeneous gamma-interval process given the underlying firing rate. 

\cite{berkes09} also propose to use copula model for capturing neural dependencies. They explore a variety of copula models for joint neural response distributions and develop an efficient maximum likelihood procedure for inference. Unlike their method, our proposed copula model in this paper is specified within a semiparametric Bayesian framework that uses Gaussian process model to obtain smooth estimates of firing rates.

Throughout this paper, we study the performance of our proposed method using simulated data and apply it to data from an experiment investigating the role of prefrontal cortical area in rats with respect to reward-seeking behavior and inhibition of reward-seeking in the absence of a rewarded outcome. In this experiment, the activity of 5-25 neurons from prefrontal cortical area was recorded. During recording, rats chose to either press or withhold presses to presented levers. Pressing lever 1 allowed the rat to acquire a sucrose reward while pressing lever 2 had no effect. (All protocols and procedures followed National Institute of Health guidelines for the care and use of laboratory animals.)

In what follows, Section \ref{gp}, we first describe our Gaussian process model for the firing rate of a single neuron. In Section \ref{twoNeurons}, we present our method for detecting co-firing (possibly after some lag time) patterns for two neurons. The extension of this method for multiple neurons is presented in Section \ref{multipleNeurons}. In Section \ref{computation}, we provide the details of our sampling algorithms. Finally, in Section \ref{discussion}, we discuss future directions.

\section{Gaussian process model of firing rates}\label{gp}
To model the underlying firing rate, we use a Gaussian process model. First we discretize time so that there is at most one spike within each time interval. Denote the response variable, $y_t$, to be a binary time series comprised of 1s (spike) and 0s (silence). The firing rate for each neuron is assumed to depend on an underlying latent variable, $u(t)$, which has a Gaussian process prior. In statistics and machine learning, Gaussian processes are widely used as priors over functions. Similar to the Gaussian distribution, a Gaussian process is defined by its mean (usually set to 0 in prior) and its covariance function $C$: $f \sim \mathcal{GP}(0, C)$. Here, the function of interest is the underlying latent variable, which is a stochastic process indexed by time $t$, $u(t)$. Hence, the covariance function is defined in terms of $t$. We use the following covariance form, which includes a wide range of smooth nonlinear functions \citep{rasmussen06, nealGP98}:
\begin{eqnarray*}
 C_{ij} & = & Cov[u(t_{i}), u(t_{j})]  \\
&=& \lambda^{2} + \eta^2 \exp [ -\rho^{2}(t_{i} - t_{j} )^2 ]  + \delta_{ij} \sigma^{2}_{\epsilon}
\end{eqnarray*}
In this setting, $\rho^{2}$ and $\eta^{2}$ control smoothness and the height of oscillations respectively. $\lambda, \eta, \rho$ and $\sigma$ are hyperparameters with their own hyperpriors. Throughout this paper, we put $N(0, 3^{2})$ prior on the log of these hyperparameters. 

We specify the spike probability, $p_t$, within time interval $t$ in terms of $u(t)$ through the following transformation:
\begin{eqnarray*}
p_t &=& \frac{1}{1+\exp [-u(t)]},
\end{eqnarray*}
As $u(t)$ increases, so does $p_t$. 

\begin{figure}[t]
\begin{center}
\centerline{\includegraphics[width=4in, height=1.7in]{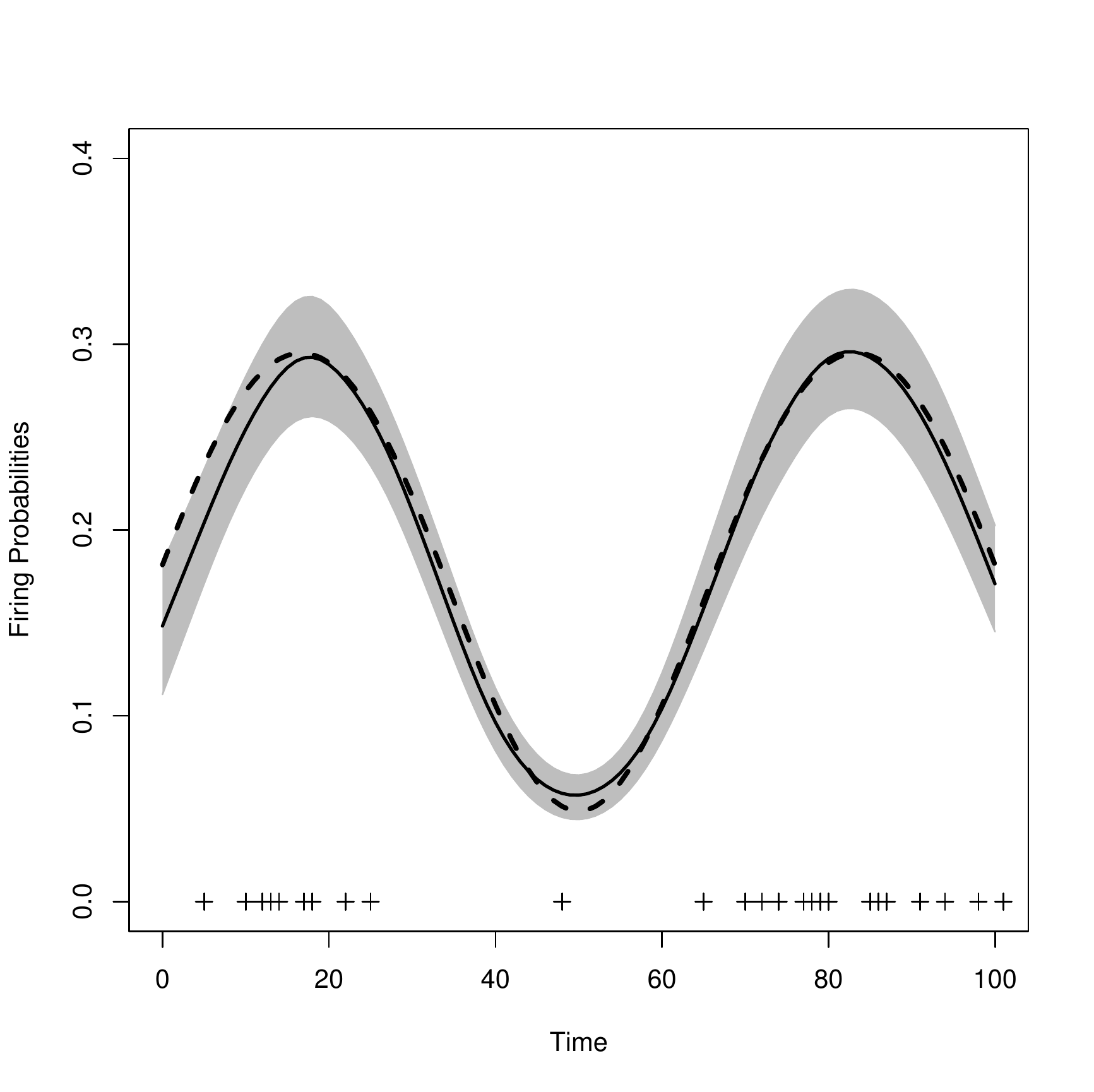}}
\caption{ An illustrative example for using a Gaussian process model for a neuron with 40 trials. The dashed line shows the true firing rate, the solid line shows the posterior expectation of the firing rate, and the gray area shows the corresponding 95\% probability interval. The plus signs on the horizontal axis represents spikes over 100 time intervals for one of the 40 trials. }\label{oneNeuron}
\vspace{-10pt}
\end{center}
\end{figure}

The prior autocorrelation imposed by this model allows the firing rate to change smoothly over time. Note that this does not mean that we believe the firing patterns over a single trial are smooth. However, over many trials, our method finds a smooth estimate of the firing rate. In posterior, of course, the data can overwhelm such prior. The dependence on prior firing patterns is through the term $(t_{i} - t_{j})$ in the covariance function. As this term decreases, the correlation between $u(t_{i})$ and $u(t_{j})$ increases. This is different from other methods \citep{kass01, kelly12} that are based on including an explicit term in the model to capture firing history. For our analysis of experimental data, we discretize the time into 5 ms intervals so there is at most one spike within each interval. Therefore, the temporal correlations in our method are on a slow time scale \citep{harrison13}. 

\begin{figure}[t]
\begin{center}
\centerline{\includegraphics[width=4.5in, height=2.5in]{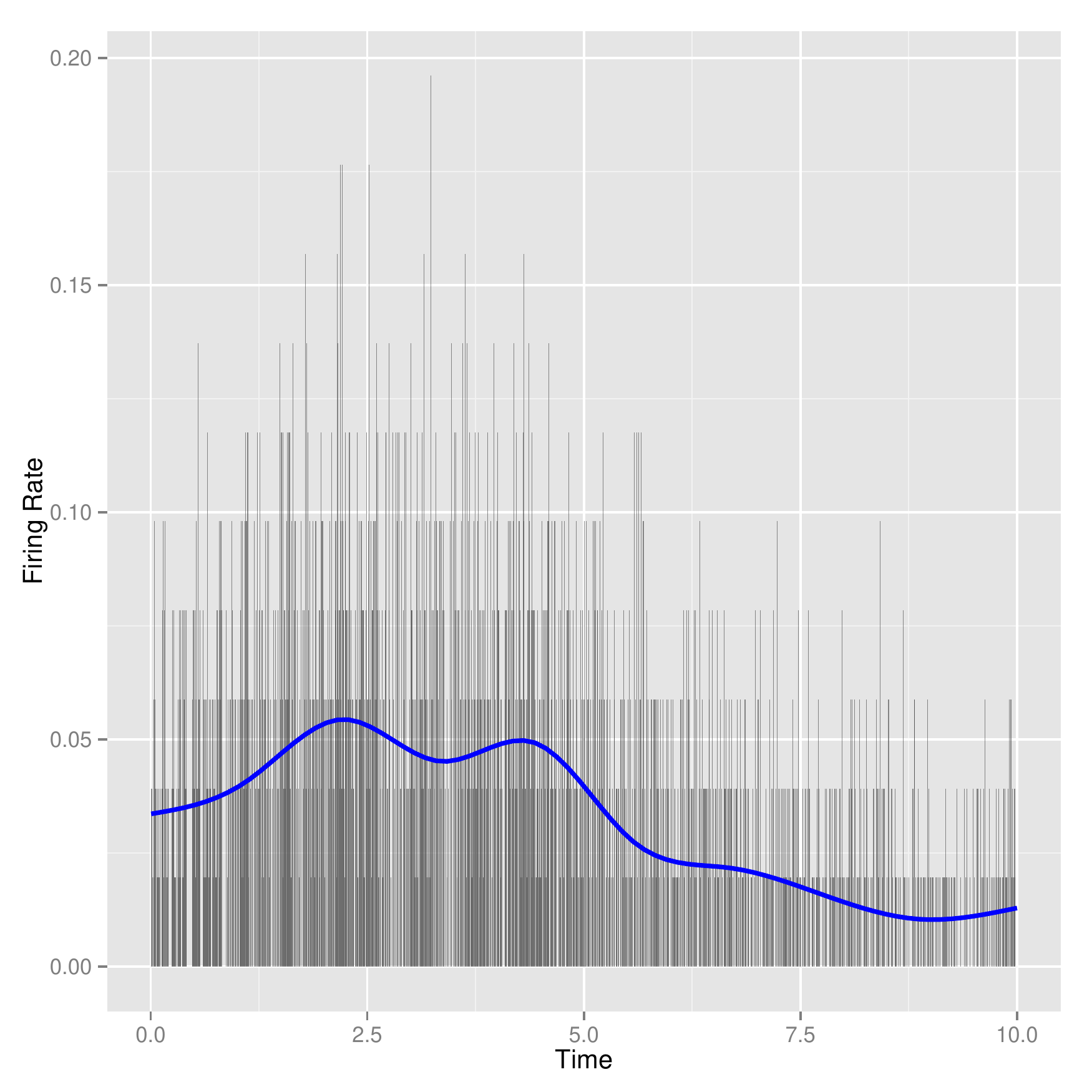}}
\caption{Using our Gaussian process model to capture the underlying firing rate of a single neuron from prefrontal cortical areas in rat's brain. There are 51 spike trains recorded over 10 seconds. The PSTH plot is generated by creating 5 ms intervals. The blue curve shows the estimated firing rate (posterior expectation).}\label{oneNeuronReal}
\vspace{-10pt}
\end{center}
\end{figure}
 
When there are $R$ trials (i.e., $R$ spike trains) for each neuron, we model the corresponding spike trains as conditionally independent given the latent variable $u(t)$. Note that we can allow for trial-to-trial variation by including a trial-specific mean parameter such that $[u(t)]^{(r)} \sim \mathcal{GP}(\mu_{r}, C)$, where $r=1, \ldots, R$, ($R=$ total number of trials or spike trains).

Figure \ref{oneNeuron} illustrates this method using 40 simulated spike trains for a single neuron. The dashed line shows the true firing rate , $p_{t} = 5(4+3\sin(3\pi t))$, for $t=0, 0.01, \ldots, 1$, the solid line shows the posterior expectation of the firing rate, and the gray area shows the corresponding 95\% probability interval. The plus signs on the horizontal axis represents spikes over 100 time intervals for one of the 40 trials. 

Figure \ref{oneNeuronReal} shows the posterior expectation of firing rate (blue curve) overlaid on the PSTH plot of a single neuron with 5 ms bin intervals from the experimental data (discussed above) recorded over 10 seconds.

\section{Modeling dependencies between two neurons}\label{twoNeurons}
Let $y_t$ and $z_t$ be binary data indicating presence or absence of spikes within time interval $t$ for two neurons. Denote $p_t$ to be the spike probability at interval $t$ for the first neuron, and $q_t$ to denote the spike probability at the same interval for the second neuron. Given the corresponding latent variables $u(t)$ and $v(t)$ with Gaussian process priors $\mathcal{GP}(0,C_u)$ and $\mathcal{GP}(0,C_v)$ respectively, we model these probabilities as $p_t = 1/\{1+\exp[-u(t)]\}$ and $q_t = 1/\{1+\exp[-v(t)]\}$.

If the the two neurons are independent, the probability of firing at the same time is $P(y_t = 1, z_t = 1) = p_t q_t$. In general, however, we can write the probability of firing simultaneously as the product of their individual probabilities multiplied by a factor, $p_tq_t\zeta$, where $\zeta$ represents the excess firing rate ($\zeta>1$) or the suppression firing rate ($\zeta<1$) due to dependence between two neurons \citep{ventura05, kelly12}. That is, $\zeta$ accounts for the excess joint spiking beyond what is explained by independence. For independent neurons, $\zeta=1$. Sometimes, the extra firing can occur after some lag time $L$. That is, in general, $P(y_t = 1, z_{t+L} = 1) = p_tq_{t+L}\zeta$ for some $L$. Therefore, the marginal and joint probabilities are
\begin{eqnarray*}
P(y_t=1|p,q,\zeta,L) &=& p_t\\
P(y_t=0|p,q,\zeta,L) &=& 1-p_t\\
P(z_{t}=1|p,q,\zeta,L) &=& q_t\\
P(z_{t}=0|p,q,\zeta,L) &=& 1-q_t\\
P(y_t=1,z_{t+L}=1|p,q,\zeta,L) &=& p_tq_{t+L}\zeta\\
P(y_t=1,z_{t+L}=0|p,q,\zeta,L) &=& p_t-p_tq_{t+L}\zeta\\
P(y_t=0,z_{t+L}=1|p,q,\zeta,L) &=& q_{t+L}-p_tq_{t+L}\zeta\\
P(y_t=0,z_{t+L}=0|p,q,\zeta,L) &=& 1-p_t- q_{t+L}+p_t q_{t+L}\zeta
\end{eqnarray*}
where
\begin{eqnarray*}
\frac{\max(p_t+q_{t+L}-1,0)}{p_tq_{t+L}} \le \zeta \le \frac{\min(p_t,q_{t+L})}{p_tq_{t+L}}
\end{eqnarray*}

In this setting, the observed data include two neurons with $R$ trials of spike trains (indexed by $r=1,2,...,R$) per neuron. Each trial runs for $S$ seconds. We discretize time into $T$ intervals (indexed by $t=1,2,...,T$) of length $S/T$ such that there are at most 1 spike in within each interval. We assume that the lag $L$ can take a finite set of values from $[-K, K]$ for some biologically meaningful $K$ and write the
likelihood function as follows:
\begin{eqnarray}\label{lagModel}
\mathcal{L}(\zeta, u, v) = \sum_{k=0}^K 1_{k=L}\prod_{r=1}^R \Big[ \prod_{t=1}^{T-L}P(y^{(r)}_{t},z^{(r)}_{t+L}) \prod_{t=T-L+1}^T P(y^{(r)}_t) \prod_{t=1}^{L}P(z^{(r)}_{t})  \Big] + \nonumber \\ 
\sum_{k=-K}^{-1} 1_{k=L}\prod_{r=1}^R \Big[\prod_{t=1}^{T+L}P(y_{t-L}^{(r)},z^{(r)}_{t}) \prod_{t=1}^{-L} P(y_{t}^{(r)}) \prod_{t=T+L+1}^{T}P(z_{t}^{(r)})  \Big]
\end{eqnarray}

We put uniform priors on $\zeta$ and $L$ over the assumed range. As mentioned above, the hyperparameters in the covariance function have weakly informative (i.e., broad) priors: we assume the log of these parameters has $N(0, 3^2)$ prior. We use Markov Chain Monte Carlo algorithms to simulate samples from the posterior distribution of model parameters given the observed spike trains. See section \ref{computation} for more details. 

\subsection{Illustrative examples}\label{sec:illust}
In this section, we use simulated data to illustrate our method. We consider three scenarios: 1) two independent neurons, 2) two dependent neurons with exact synchrony ($L=0$), and 3) Two dependent neurons with lagged co-firing. In each scenario, we assume a time-varying firing rate for each neuron and simulate 40 spike trains given the underlying firing rate. For independent neurons, we set $\zeta=1$, whereas $\zeta > 1$ for dependent neurons.

\begin{figure}[!t]
\vskip 0.2in
\begin{center}
\centerline{\includegraphics[width=1.8in, height=1.7in]{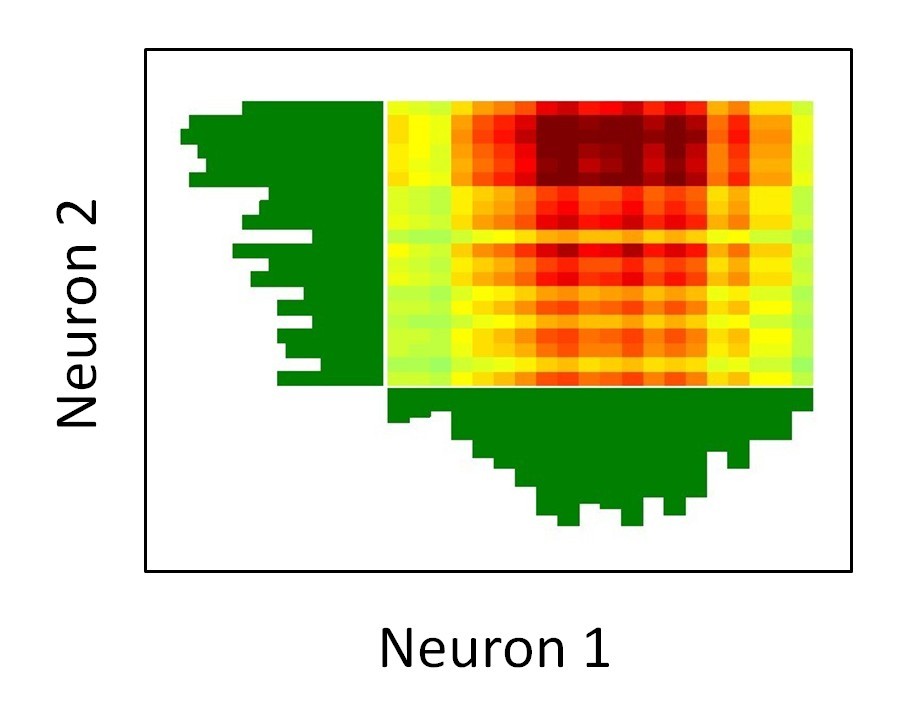}\hspace{30pt}  \includegraphics[width=2.7in, height=1.85in]{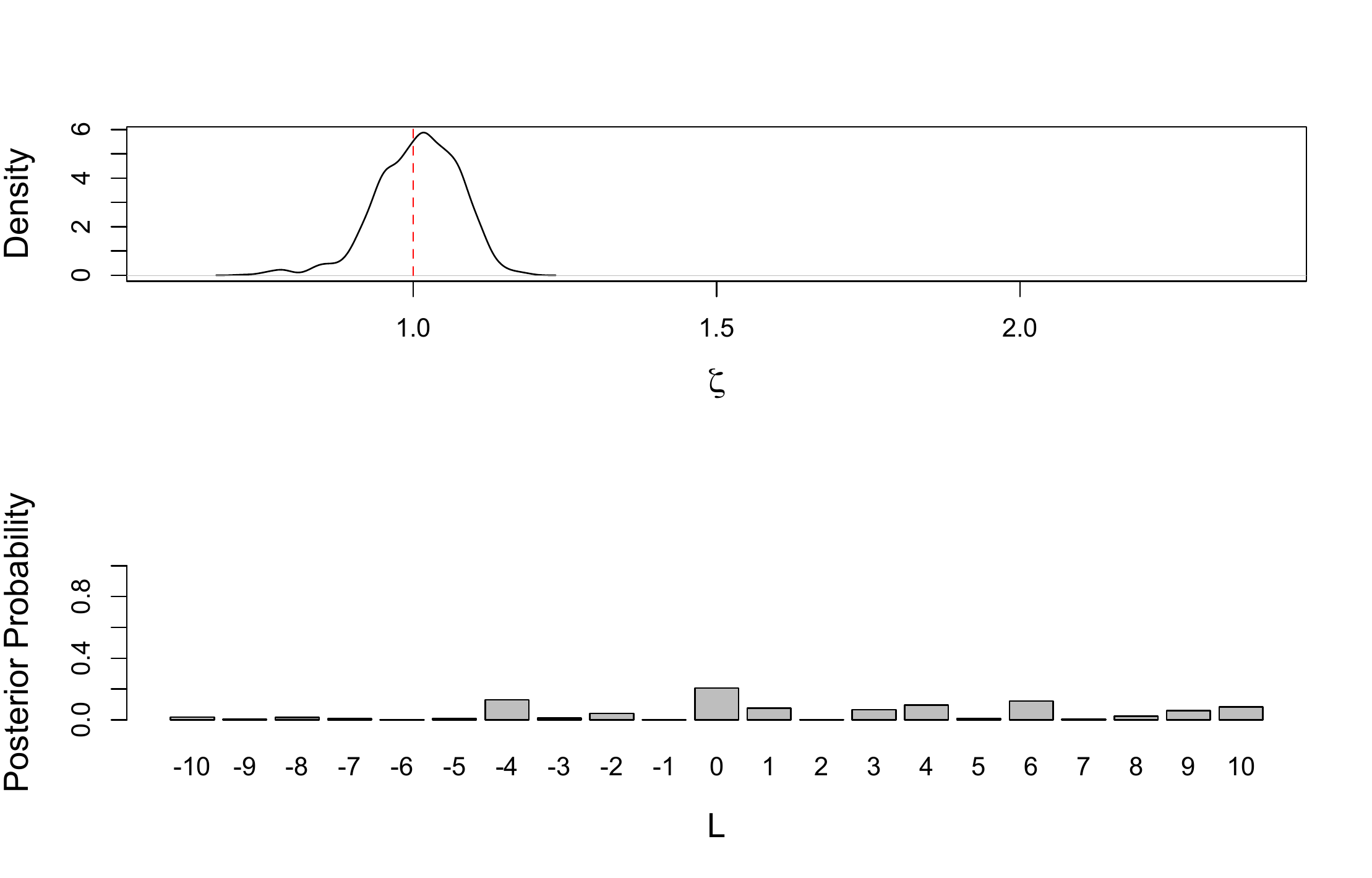}}
\vspace{-15pt}
\caption{{\bf{Two independent neurons}}-- The left panel shows the corresponding Joint Peri-Stimulus Time Histogram (JPSTH). The right panel shows the posterior distributions of $\zeta$ and $L$. Darker cells represent higher frequencies. }\label{twoIndependentNeurons}
\vspace{20pt}
\centerline{\includegraphics[width=1.8in, height=1.7in]{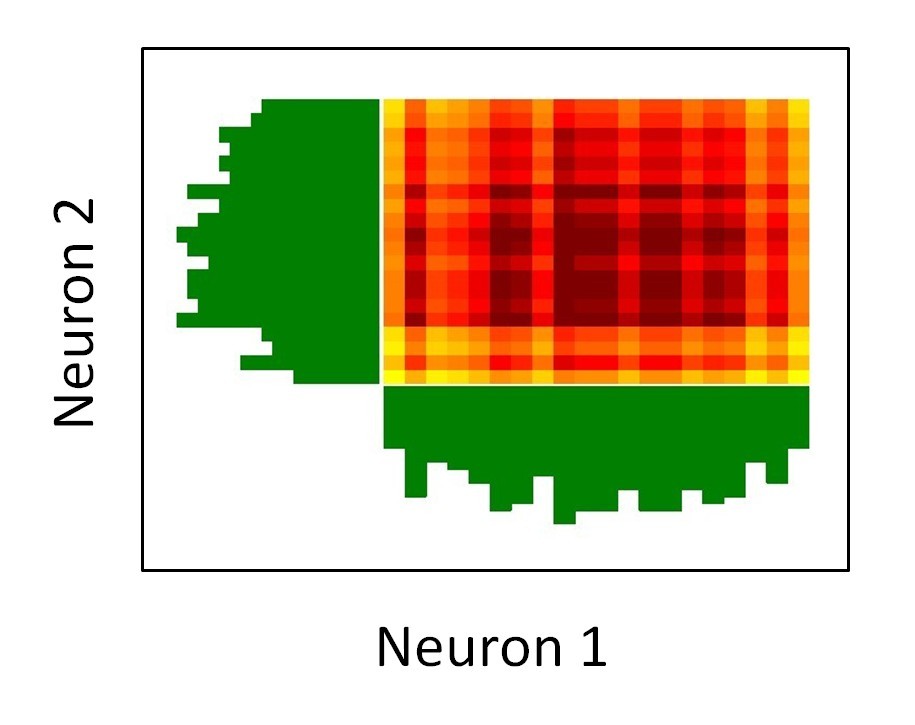} \hspace{30pt}  \includegraphics[width=2.7in, height=1.85in]{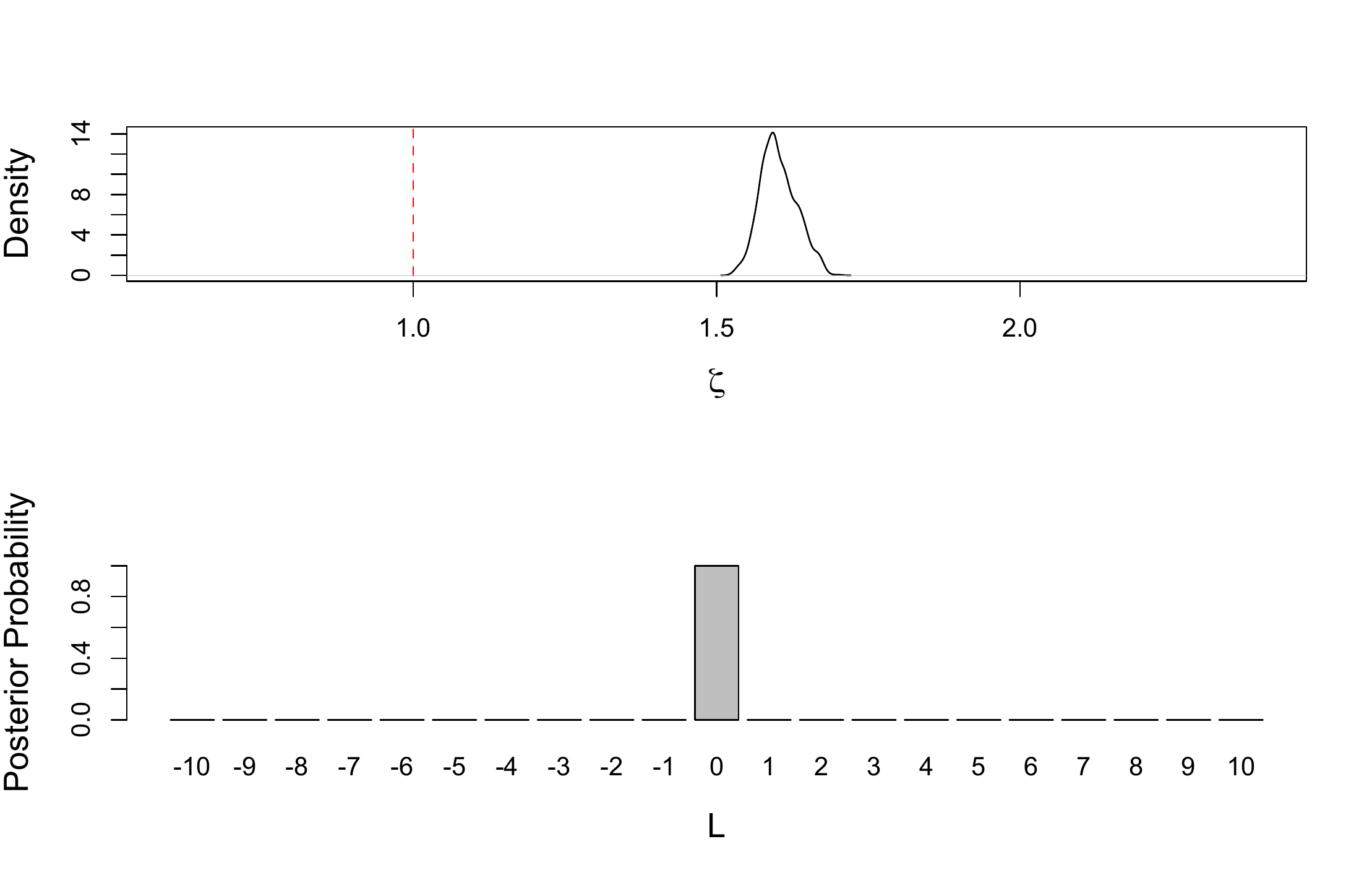}}
\vspace{-15pt}
\caption{{\bf{Two dependent neurons in exact synchrony}}-- The left panel show the frequency of spikes over time. The right panel shows the posterior distribution of $\zeta$ and $L$. Darker cells represent higher frequencies. }\label{twoDependentNeuronsNoLag}
\vspace{20pt}
\centerline{\includegraphics[width=1.8in, height=1.7in]{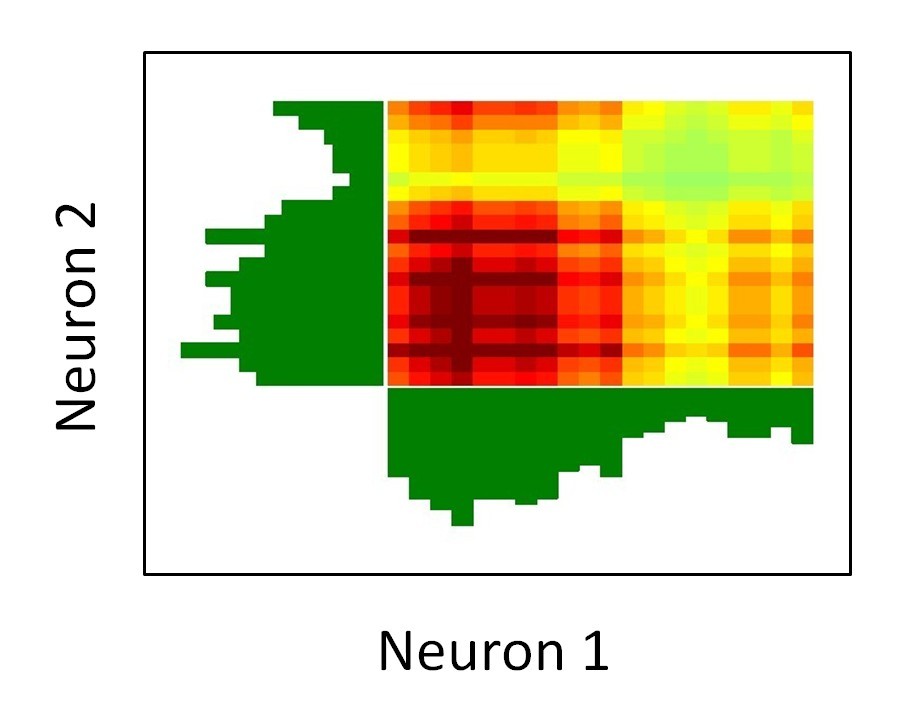} \hspace{30pt}  \includegraphics[width=2.7in, height=1.85in]{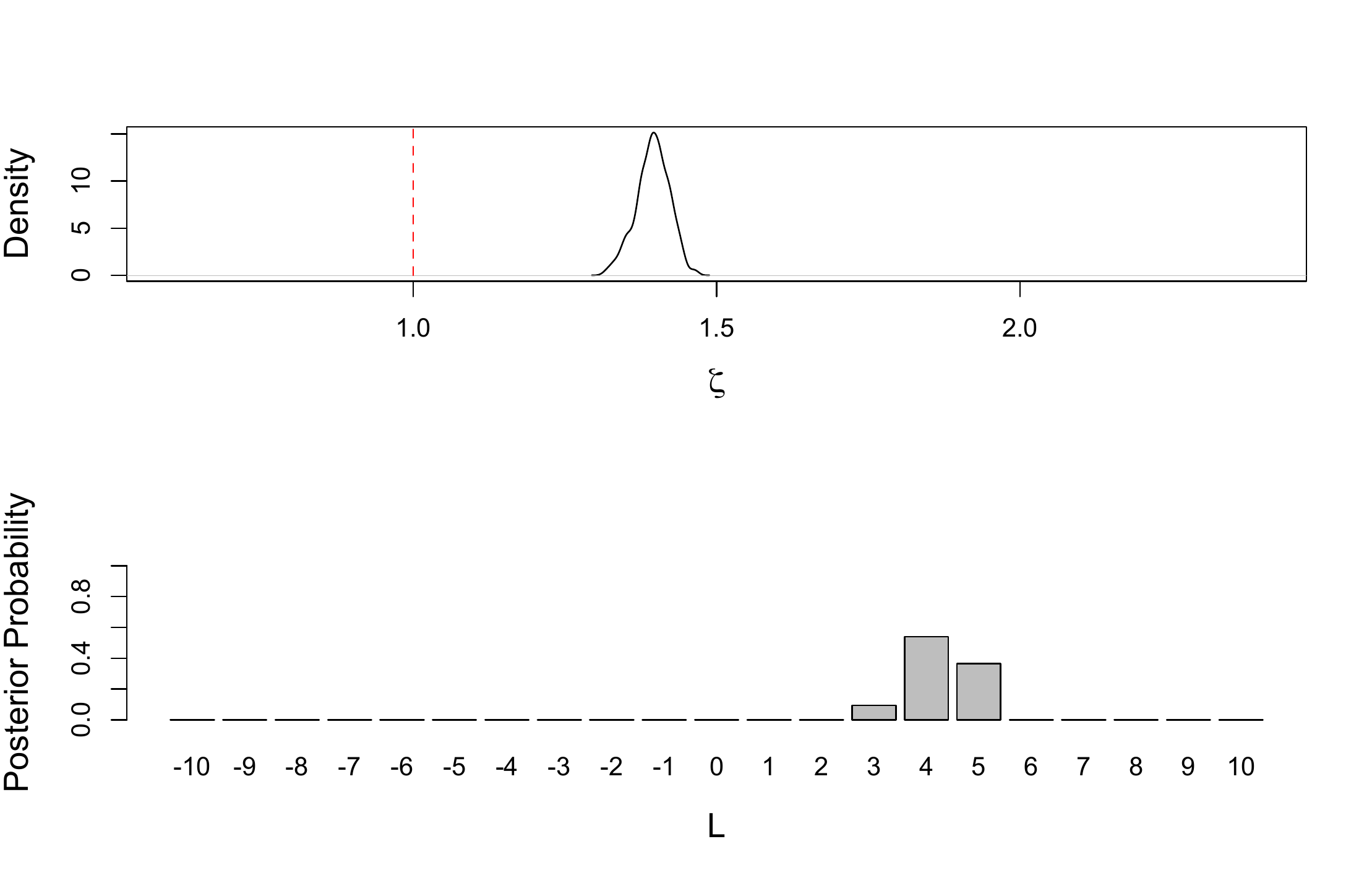}}
\vspace{-15pt}
   \caption{{\bf{Two dependent neurons in lagged synchrony}}-- The lag values are set to 3, 4, or 5 with probabilities 0.2, 0.5, and 0.3 respectively. The left panel show the frequency of spikes over time. The right panel shows the posterior distribution of $\zeta$ and $L$. Darker cells represent higher frequencies. }\label{twoDependentNeuronsWithLag}
    \end{center}
\vspace{20pt}
\end{figure}

\subparagraph{Two independent neurons.}
In the first scenario, we consider two independent neurons ($\zeta=1$). We simulate the spike trains according to our model. The firing probability at time $t$ is set to $0.25 - 0.1\cos(2\pi t)$ for the first neuron and to $0.15 + 0.2t$ for the second neuron. For each neuron, we generated 40 trials of spike trains and divided each trial into 100 time intervals. The left panel of Figure \ref{twoIndependentNeurons} shows the corresponding Joint Peristimulus Time Histogram (JPSTH). Each cell represents the joint frequency of spikes (darker cells represent higher frequencies) for the two neurons at given times. The marginal distributions of spikes, i.e., Peristimulus Time Histogram (PSTH), for the first neuron is shown along the horizontal axis. The second neuron's PSTH is shown along the vertical axis. The right panel of Figure \ref{twoIndependentNeurons} shows the posterior distributions of $\zeta$ and $L$. For this example, the posterior distribution of $\zeta$ is concentrated around 1 with median and 95\% posterior probability interval equal to 1.01 and [0.85,1.12] respectively. This would strongly suggest that the two neurons are independent as expected. Further, the posterior probabilities of all lag values from -10 to 10 are quite small.

\subparagraph{Two exact synchronous neurons.} For our next example, we simulate data for two dependent neurons with synchrony (i.e., $L=0$) and we set $\zeta=1.6$. That is, the probability of co-firing at the same time is 60\% higher than that of independent neurons. As before, for each neuron we generate 40 trials of spike trains each discretized into 100 time bins. In this case, the firing probabilities at time $t$ for the two neurons are $0.25 - 0.1\cos(2\pi t)$. Figure \ref{twoDependentNeuronsNoLag} shows their corresponding JPSTH along with the posterior distributions of $\zeta$ and $L$. The posterior median for $\zeta$ is $1.598$ and the 95\% posterior probability interval is [1.548,1.666]. Therefore, $\zeta$ identifies the two neurons in exact synchrony with excess co-firing rate than what is expected by independence. Further, the posterior distribution of $L$ shows that the two neurons are in exact synchrony.

\subparagraph{Two dependent neurons with lagged co-firing.}
Similar to the previous example, we set the probability of co-firing to 60\% higher than what we obtain by the independence assumption. Similar to the previous two simulations, we generate 40 trials of spike trains each discretized into 100 time bins. The firing probabilities of the first neurons at time $t$ is set to $0.25 + 0.1\sin(2\pi t)$. The second neuron has the same firing probability but at time $t+L$. For different trials, we randomly set $L$ to 3, 4, or 5 with probabilities 0.2, 0.5, and 0.3 respectively. Figure \ref{twoDependentNeuronsWithLag} shows JPSTH along with the posterior distributions of $\zeta$ and $L$. As before, the posterior distribution of $\zeta$ can be used to detect the relationship between the two neurons. For this example, the posterior median and 95\% posterior interval for $\zeta$ are 1.39 and [1.33,1.44] respectively. Also, our method could identify the three lag values correctly.

\subsection{Power analysis}\label{experiments}
\begin{figure}[t]
    \begin{center}
    \subfigure{\includegraphics[width=65mm,height=40mm]{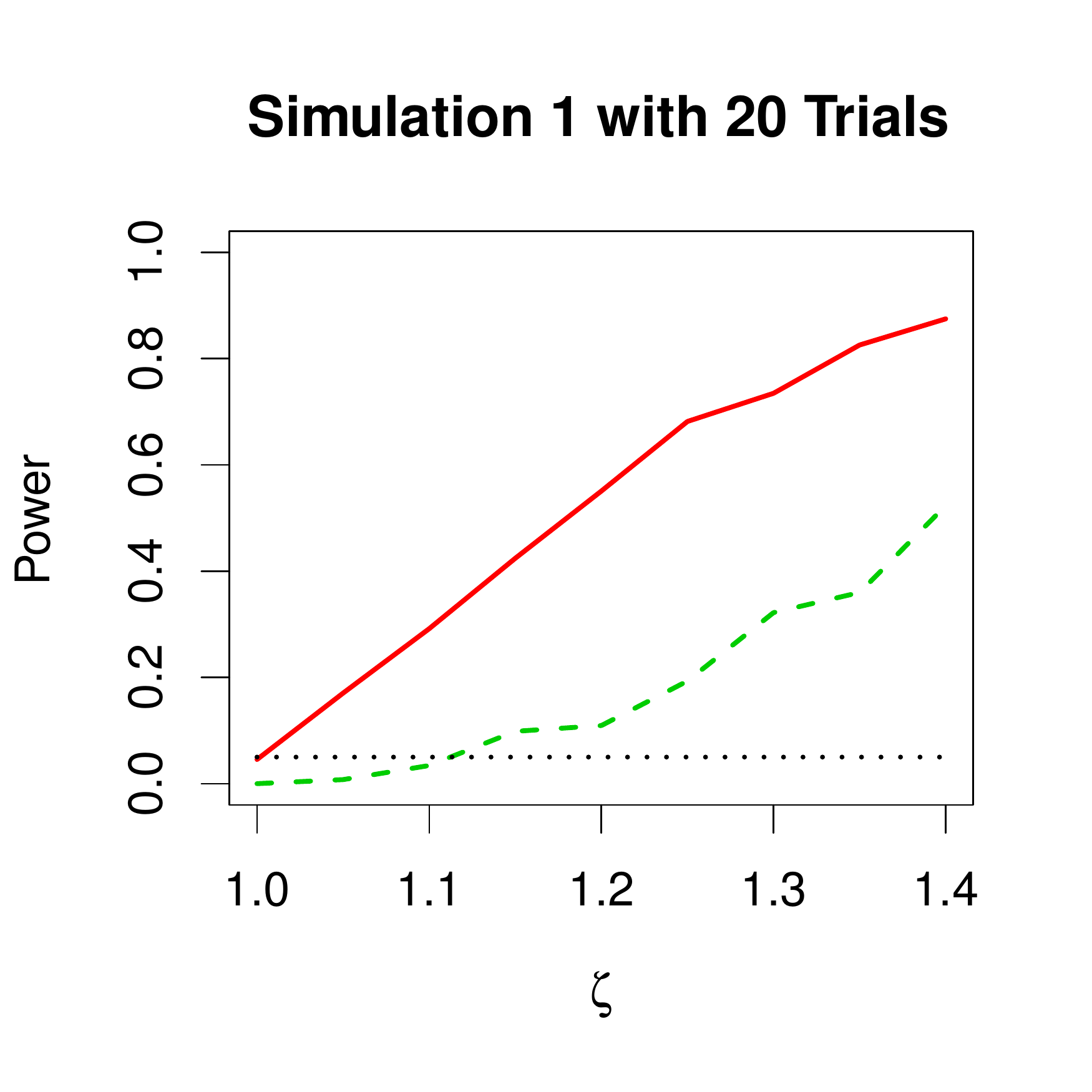}}
    \subfigure{\includegraphics[width=65mm,height=40mm]{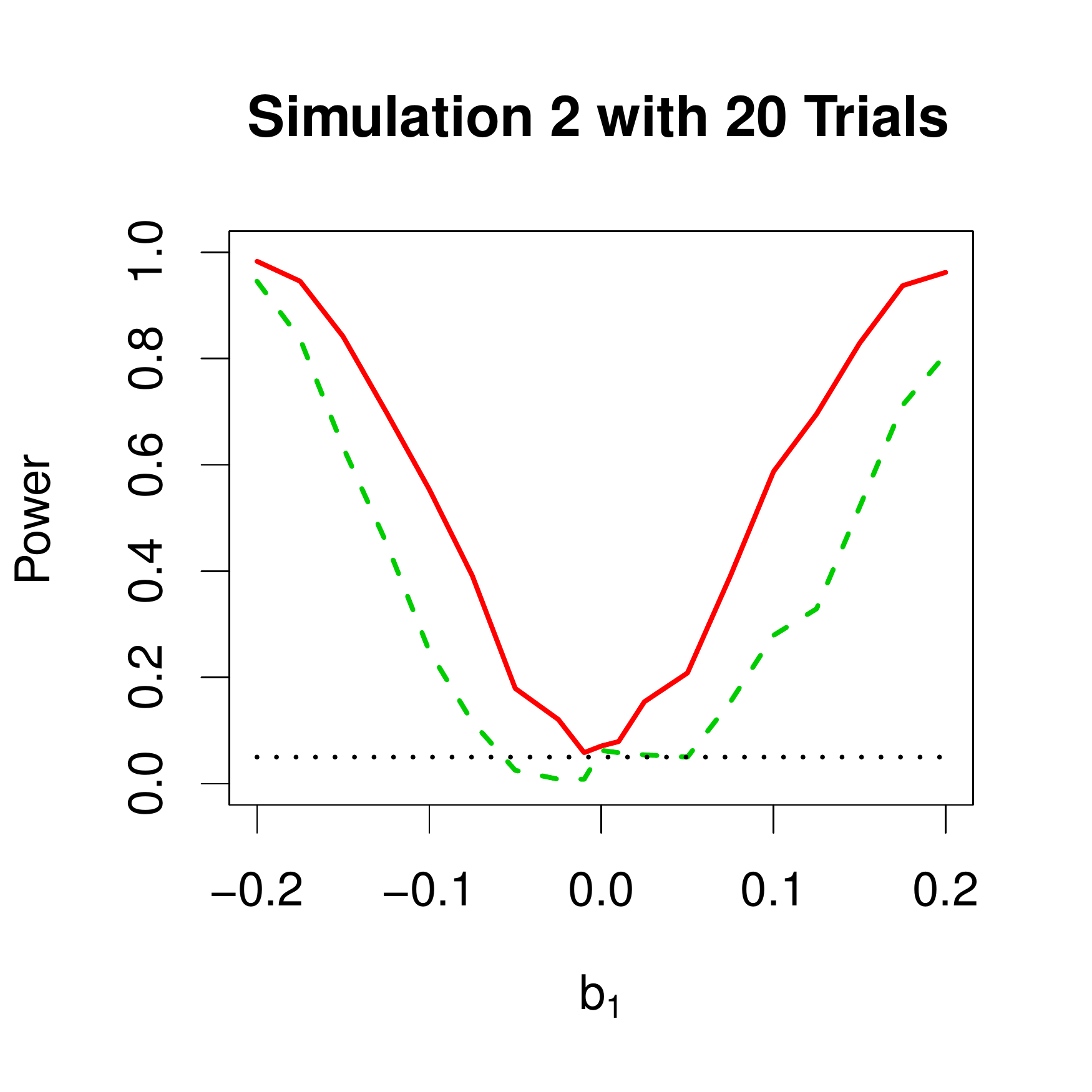}}\\
    \subfigure{\includegraphics[width=65mm,height=40mm]{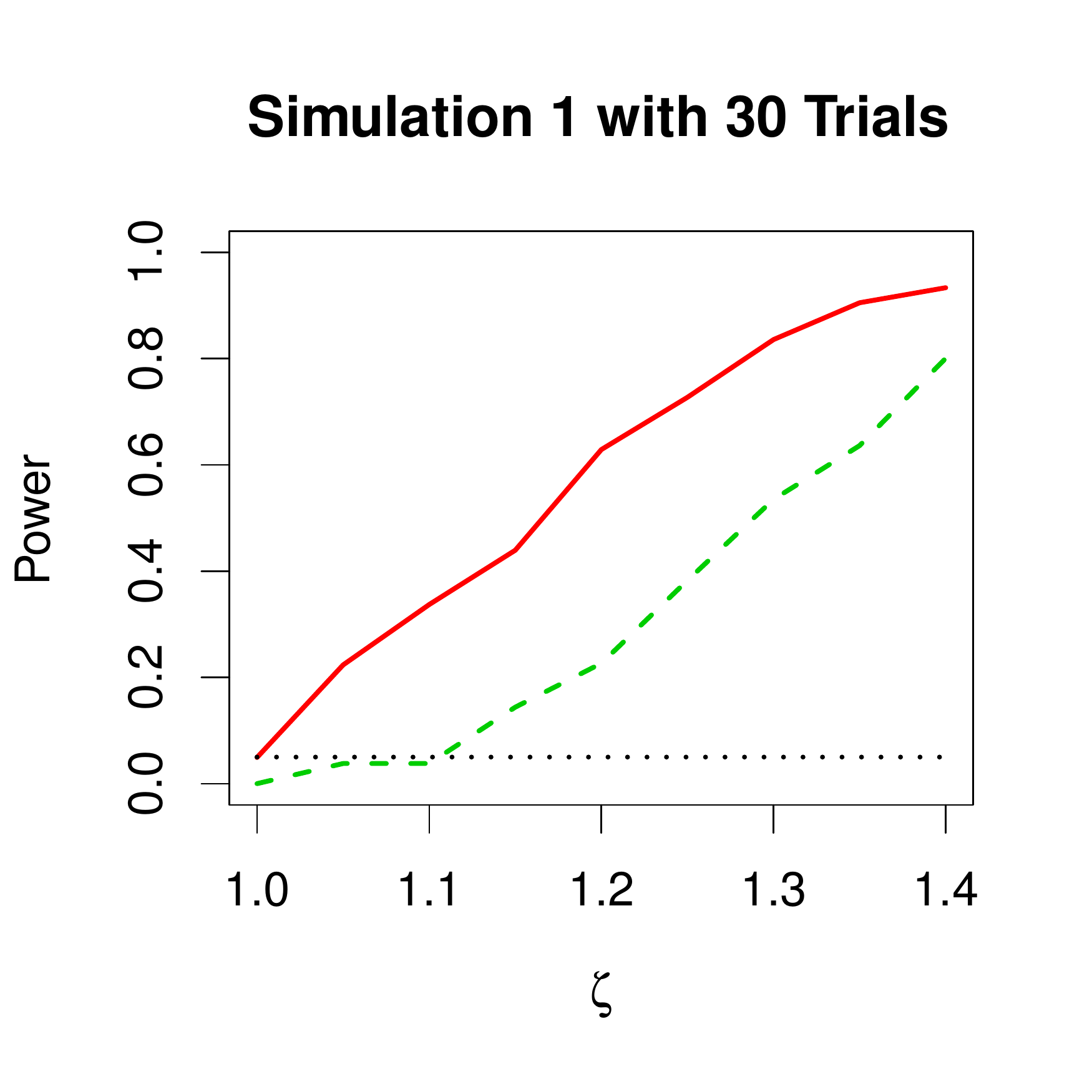}}
    \subfigure{\includegraphics[width=65mm,height=40mm]{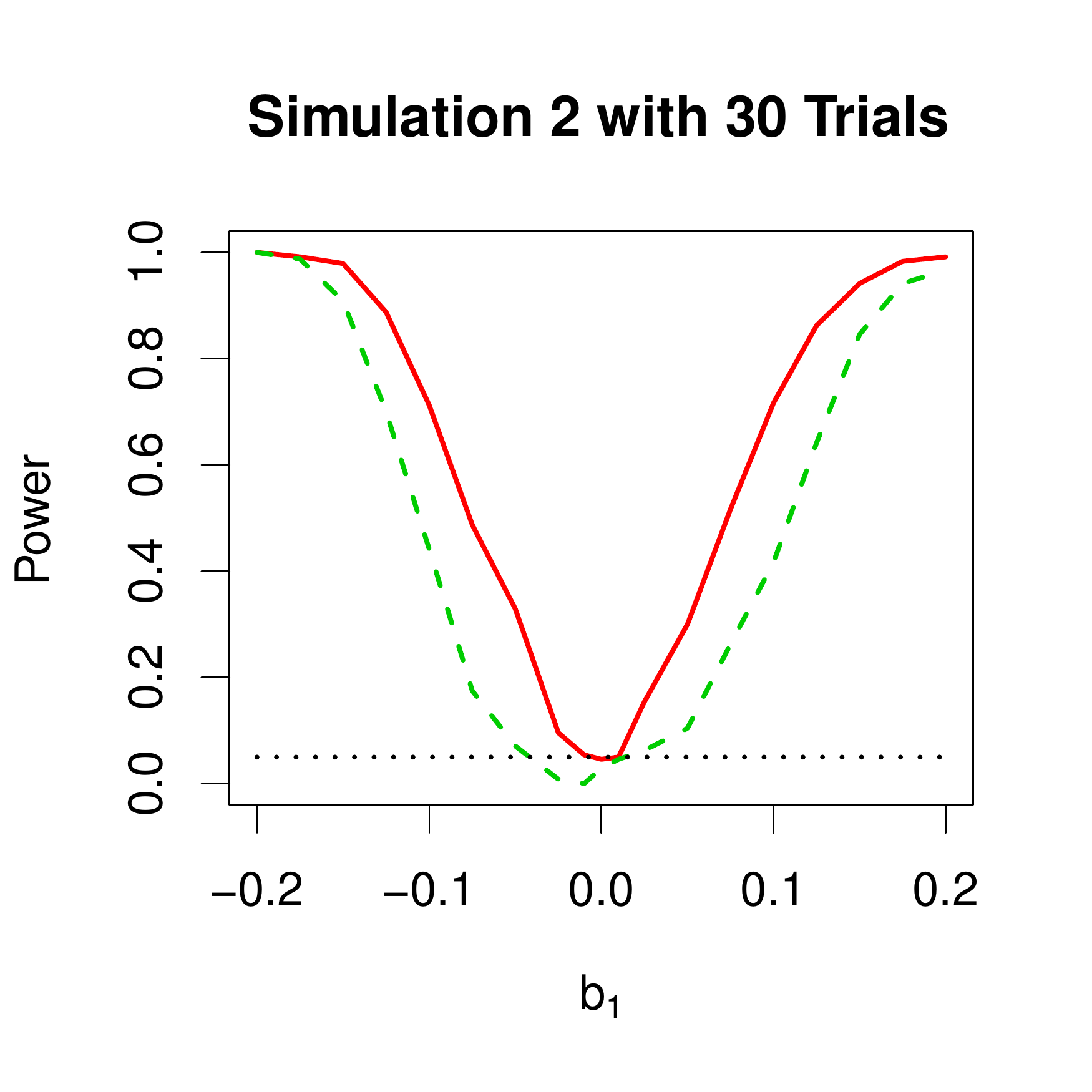}}\\
    \subfigure{\includegraphics[width=65mm,height=40mm]{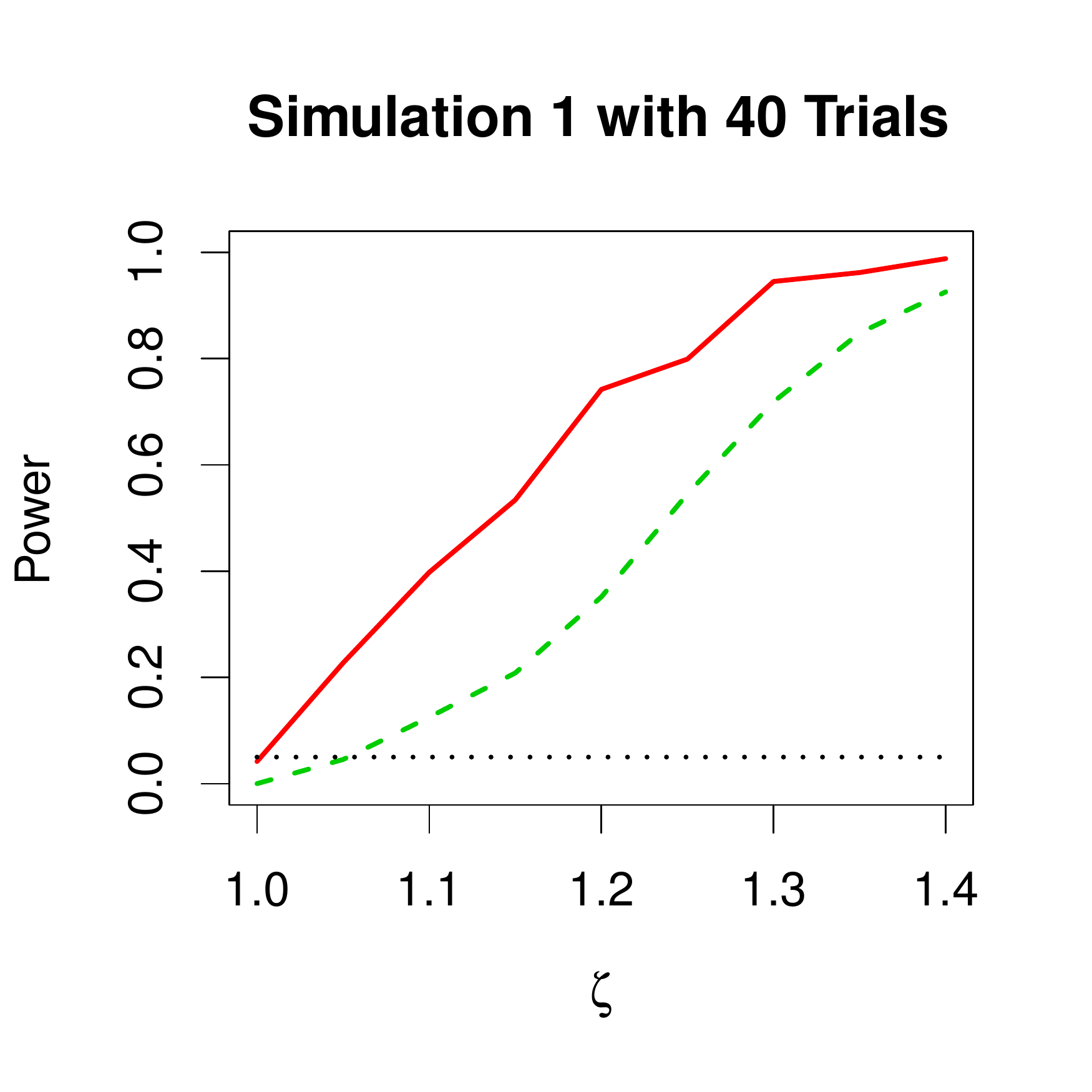}}
    \subfigure{\includegraphics[width=65mm,height=40mm]{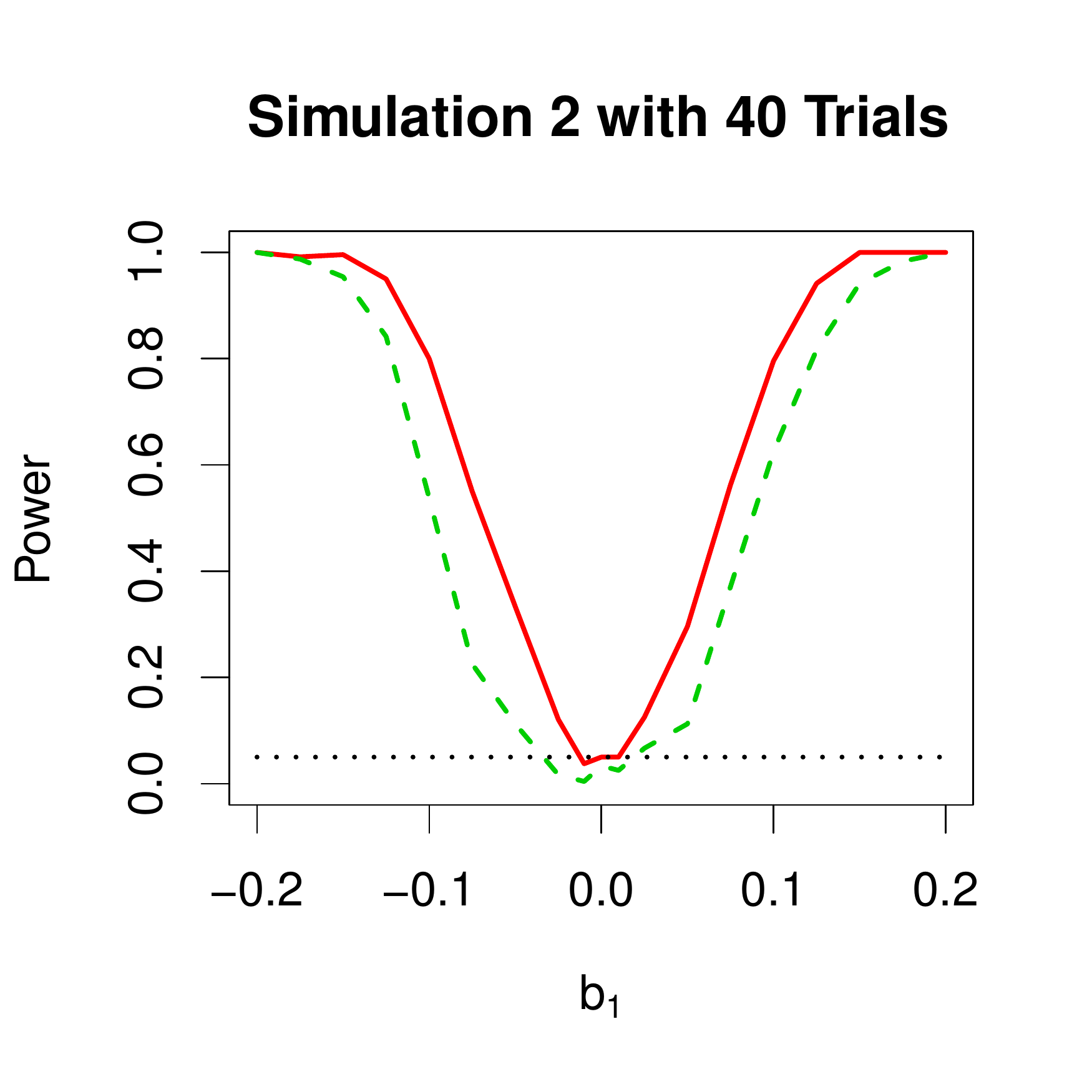}}
    \vspace{-15pt}
    \caption{ {\bf Power analysis}-- Comparing our proposed method (solid curves) to the method of \cite{kass11} (dashed curves) based on statistical power using two simulation studies. Here the dotted lines indicate the 0.05 level.}\label{power}
    \end{center}
\end{figure}

Next, we evaluate the performance of our proposed approach. More specifically, we compare our approach to the method of \cite{kass11} in terms of statistical power for detecting synchronous neurons. To be precise, given the true value of $\zeta$, we compare the ratio of correctly identifying synchrony between two neurons over a large number of simulated pairs of spike trains. In their approach, \cite{kass11} find the marginal firing rate of each neuron using natural cubic splines and then evaluate the amount of excess joint spiking using the bootstrap method. Therefore, for our first simulation study, we generate datasets that conform with the underlying assumptions of both methods. More specifically, we first set the marginal firing rates to $p_{t} = q_{t} = 0.2-0.1\cos(12\pi t)$, and then generate the spike trains for the two neurons given $\zeta$ (i.e., excess joint firing rate). The left panel of Figure \ref{power} compares the two methods in terms of statistical power for different values of $\zeta$ and different number of trials (20, 30, and 40) each with 20 time intervals. For each simulation setting, we generate 240 datasets. In our method, we call the relationship between two neurons significant if the corresponding 95\% posterior probability does not include 1. For the method proposed by \cite{kass11}, we use the 95\% bootstrap confidence intervals instead. As we can see, our method (solid curve) has substantially higher power compared to the method of \cite{kass11} (dashed curve). Additionally, our method correctly achieves 0.05 level (dotted line) when $\zeta = 1$ (i.e., the two neurons are independent).

For our second simulation, we generate datasets that do not conform with the underlying assumptions of the two methods. Let $Y=(y_1, \ldots ,y_T)$ and $Z=(z_1, \ldots ,z_T)$ denote the spike trains for two neurons. We first simulate $y_t$, i.e., absence or presence of spikes for the first neuron at time $t$, from Bernoulli$(p_{t})$, where $p_{t} = 0.25 - 0.1\cos(12\pi t)$ for $t \in [0, 0.2]$. Then, we simulate $z_t$ for the second neuron from Bernoulli$(b_0+b_1 y_t)$ for given values of $b_{0}$ and $b_{1}$. We set $b_0$ (i.e., the baseline probability of firing for the second neuron) to 0.2. When $b_1=0$, the two neurons are independent. Positive values of $b_{1}$ leads to higher rates of co-firing between the two neurons. When $b_{1}$ is negative, the first neuron has an inhibitory effect on the second neuron. For given values of $b_1$ and number of trials (20, 30, and 40), we generate 240 datasets where each trial has 20 time intervals. The right panel of Figure \ref{power} compares the two methods in terms of statistical power under different settings. As before, our method (solid curves) has higher statistical power compared to the method of \cite{kass11} (dashed curves).

\subsection{Sensitivity analysis for trial-to-trial variability}\label{sensitivity}
\begin{figure}[t]
  \begin{center}
  \includegraphics[width=80mm,height=60mm]{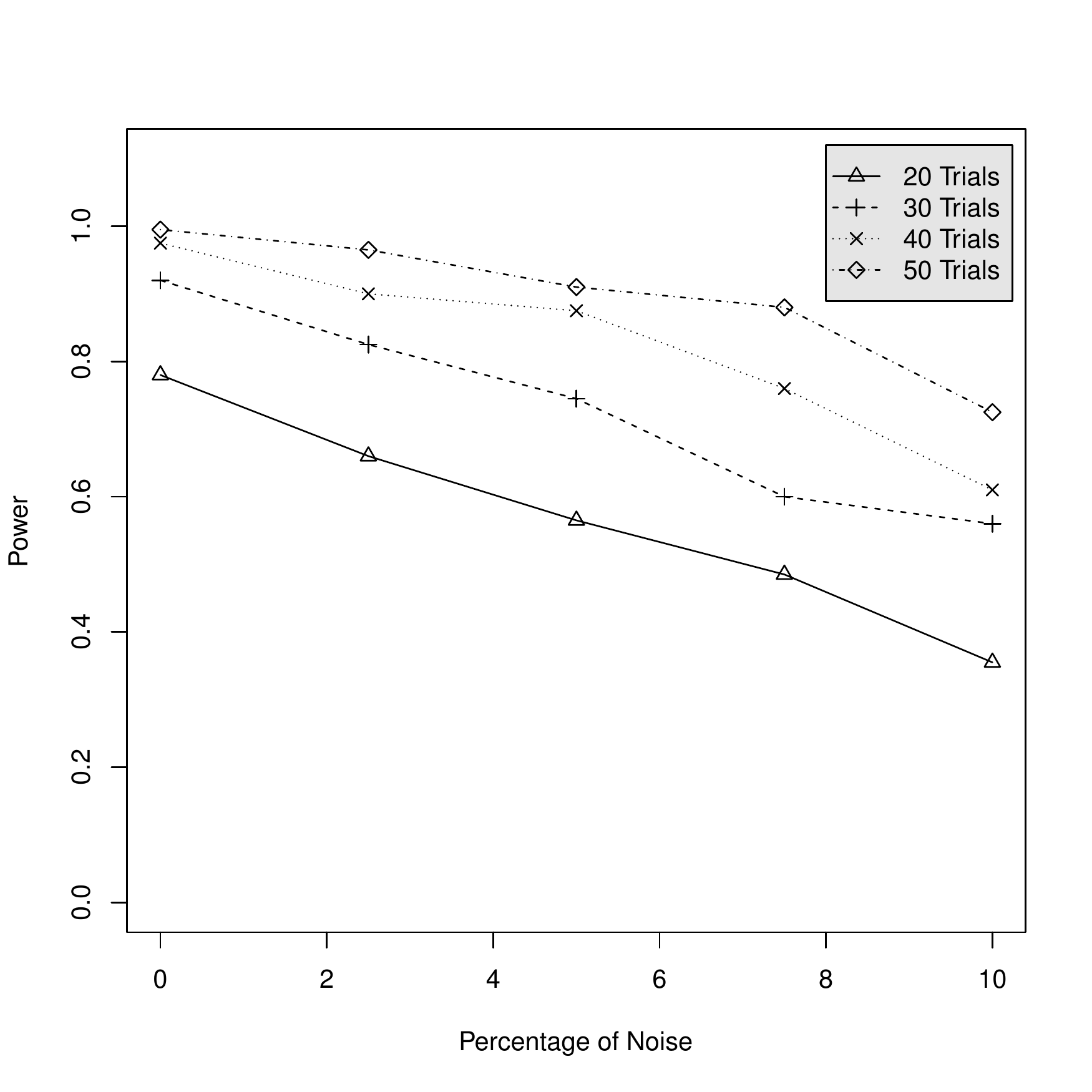}
  \caption{ {\bf Sensitive analysis for trial-to-trial variability}-- Comparing power for varying number of trials and noise rate (i.e., fraction of time bins in a trial flipped from zero to one or from one to zero).}\label{sens}
  \end{center}
\end{figure}

As mentioned above, our method can be easily extended to allow for trial-to-trial variability. To examine how such variability can affect our current model, we conduct a sensitivity analysis. Similar to the procedure discussed in the previous section, we start by setting the underlying firing probabilities to $p_{t}=0.4+0.1\cos(12t)$ and $\zeta=1.2$. For each simulated dataset, we set the number of trials to 20, 30, 40, and 50. We found that shifting the firing rate of each trial by a uniformly sampled constant around the true firing rate does not substantially affect our method's power since the Gaussian process model is still capable of estimating the underlying firing rate by averaging over trials. However, adding independent random noise to each trial (i.e., flipping a fraction of time bins from zero to one or from one to zero) could affect performance, especially if the noise rate (i.e., proportion of flips) is high and the number of trials is low. Figure \ref{sens} shows the power for different number of trials and varying noise rate from 0 to 10\%. As we can see, the power of our method drops slowly as the percentage of noise increases. The drop is more substantial when the number of trials is small (i.e., 20). However, for a reasonable number of trials (e.g., 40 or 50) and a reasonable noise rate (e.g., about 5\%) the drop in power is quite small. 
 
\subsection{Results for experimental data}\label{sec:illust}

We now use our method for analyzing a pair of neurons selected from the experiment discussed in the introduction. (We will apply our method to multiple neurons in the next section.) Although we applied our method to several pairs with different patterns, for brevity we present the results for two pair of neurons; for one pair, the relationship changes under different scenarios; for the other pair, the relationship remain the same under the two scenarios. Our data include 51 spike trains for each neuron under different scenario (rewarded vs. non-rewarded). Each trail runs for 10 seconds. We discretize the time into 5 ms intervals.

\subparagraph{Case 1: Two neurons with synchrony under both scenarios}

\begin{figure}[!t]
  \begin{center}
  \subfigure[]{\includegraphics[width=2.7in,height=1.4in]{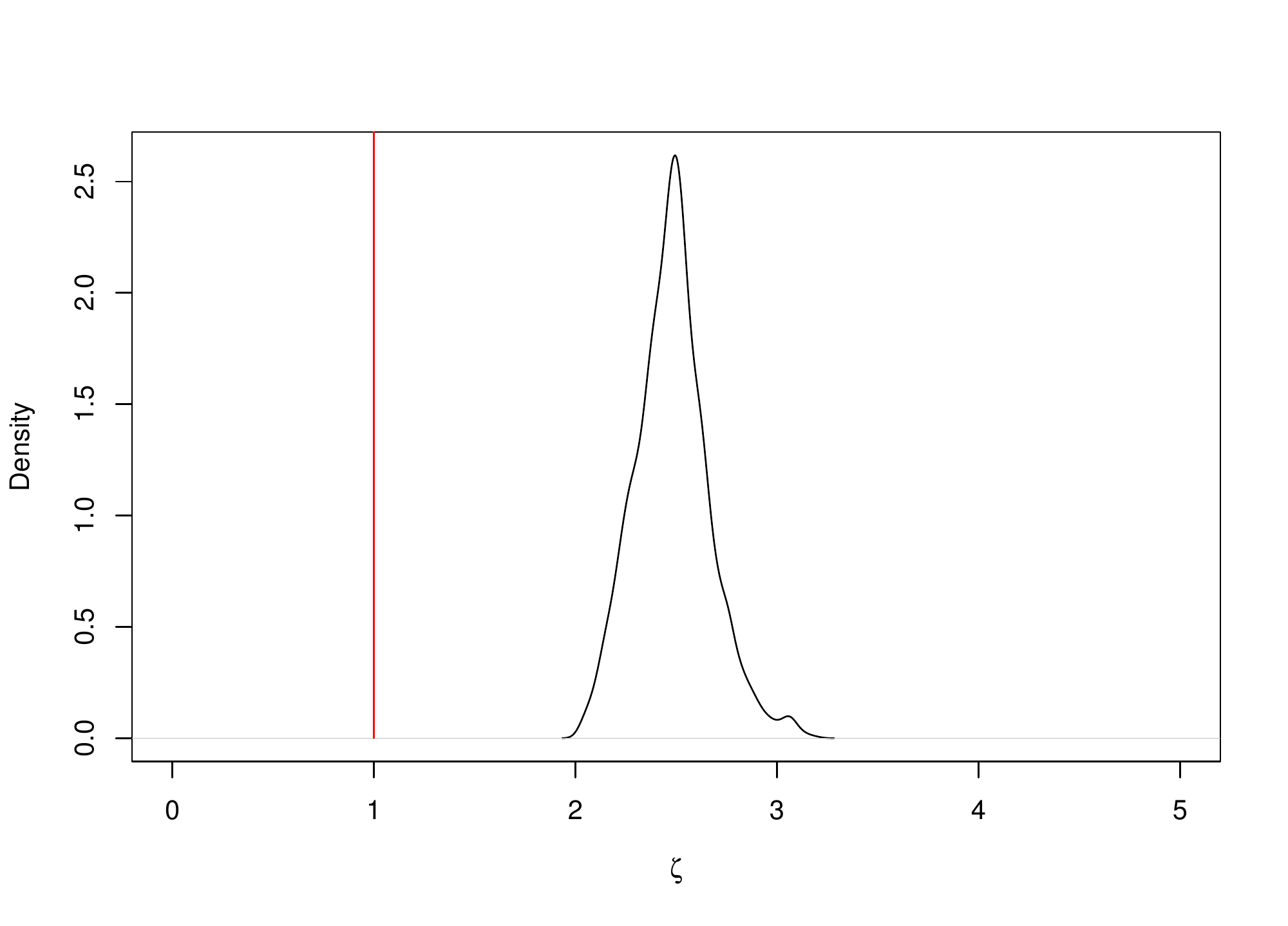}}\label{1a}
  \subfigure[]{\includegraphics[width=2.7in,height=1.4in]{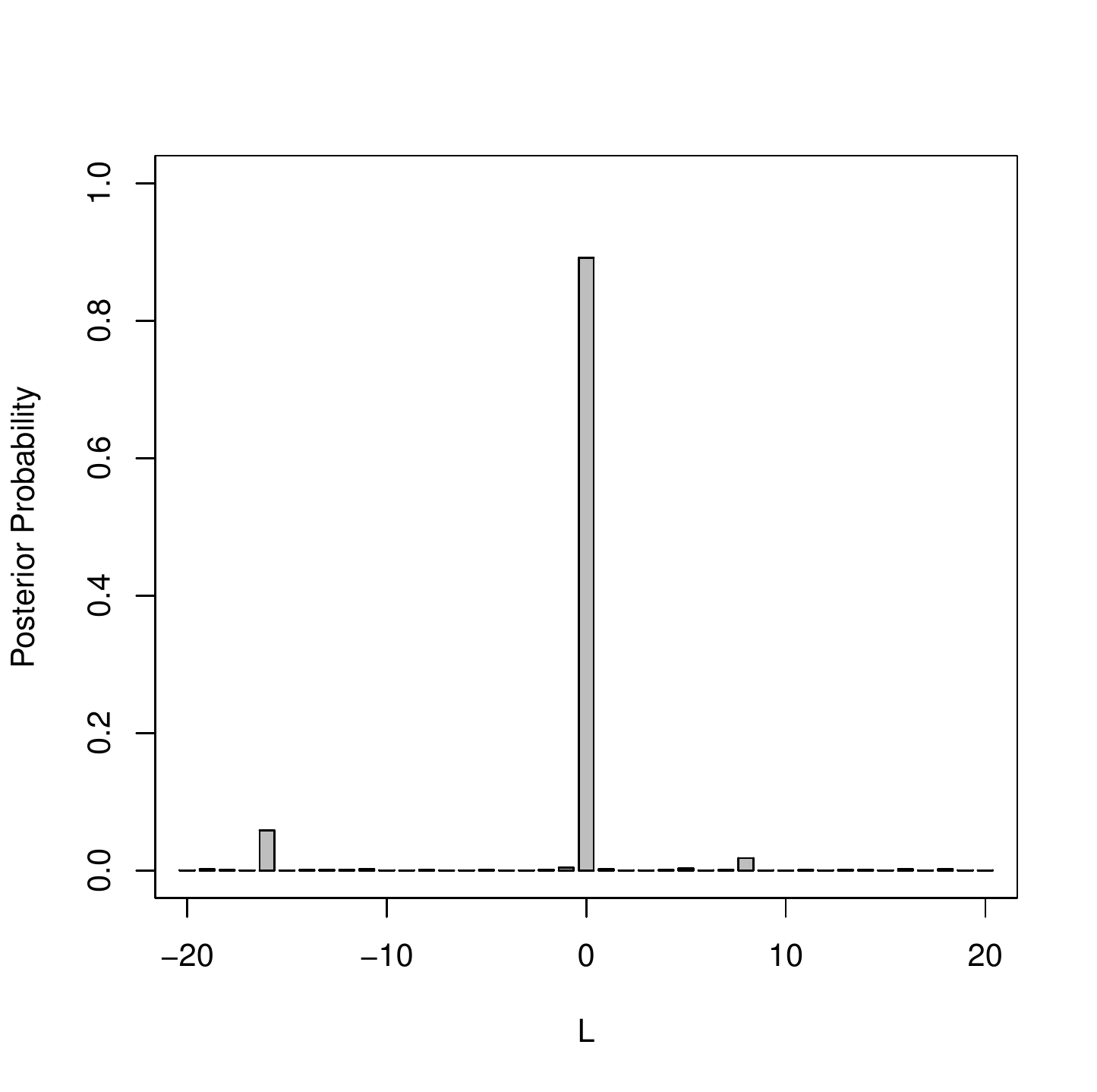}}\label{1b}

\subfigure[]{\includegraphics[width=1.8in, height=1.4in]{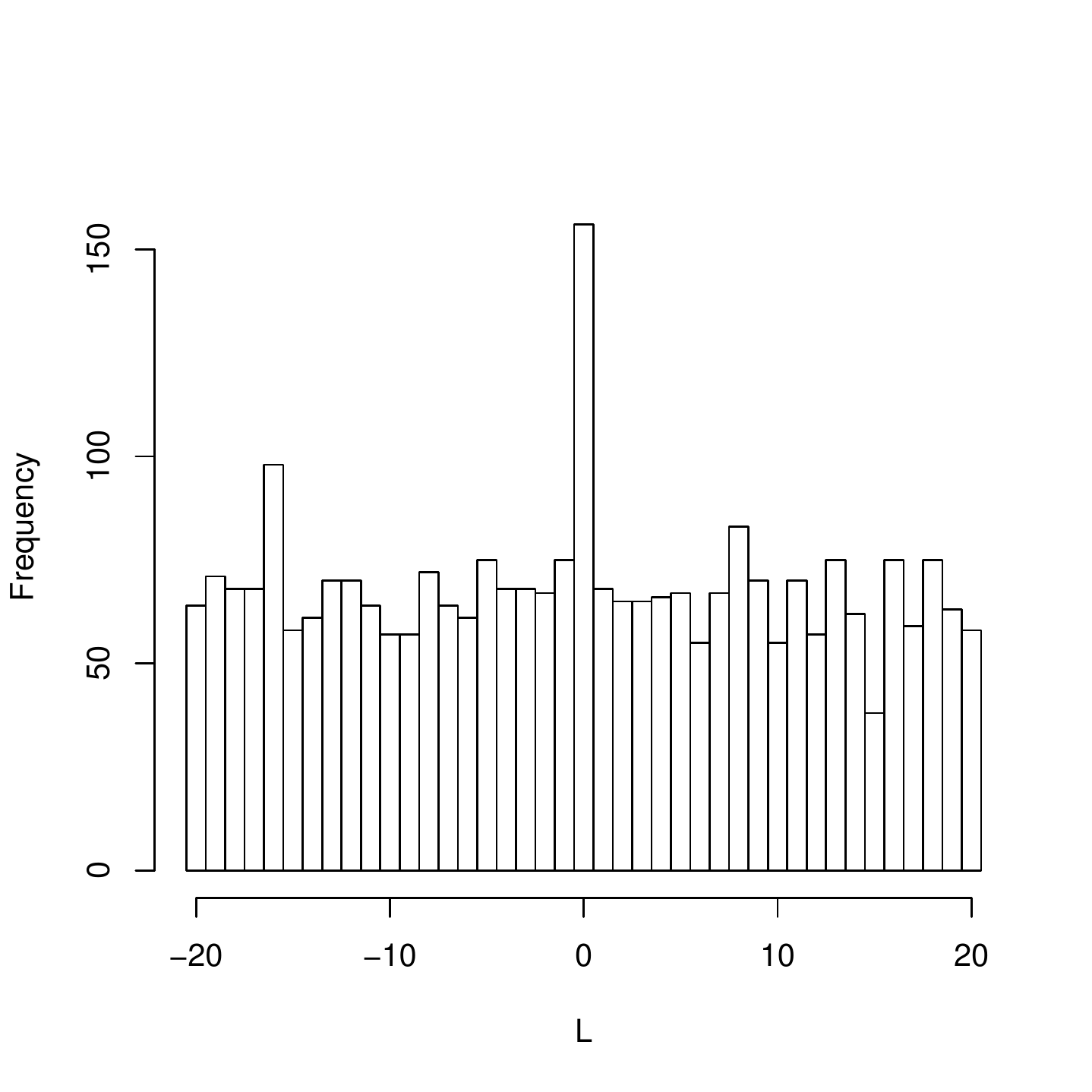}} \subfigure[]{\includegraphics[width=1.8in, height=1.4in]{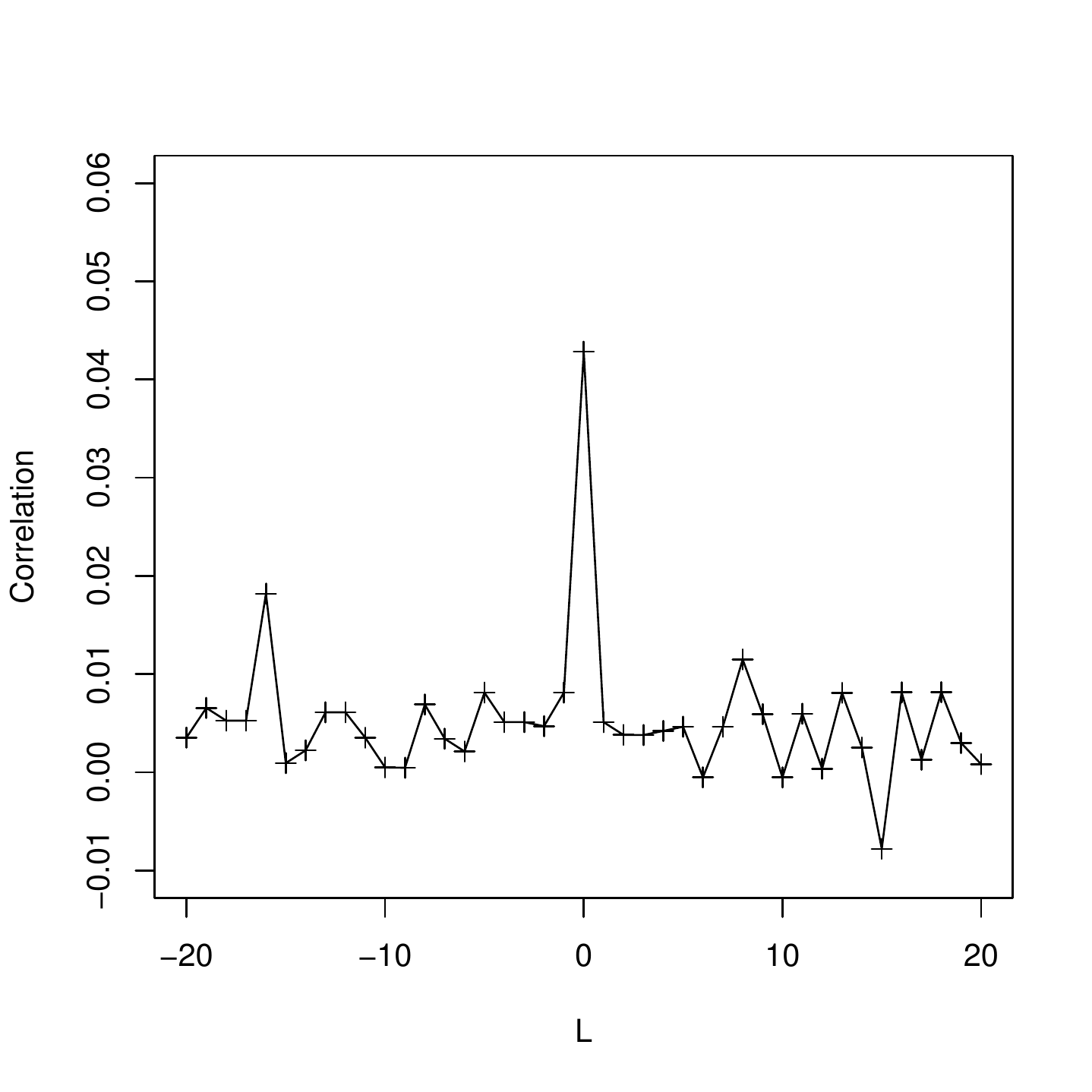}} \subfigure[]{\includegraphics[width=1.8in, height=1.4in]{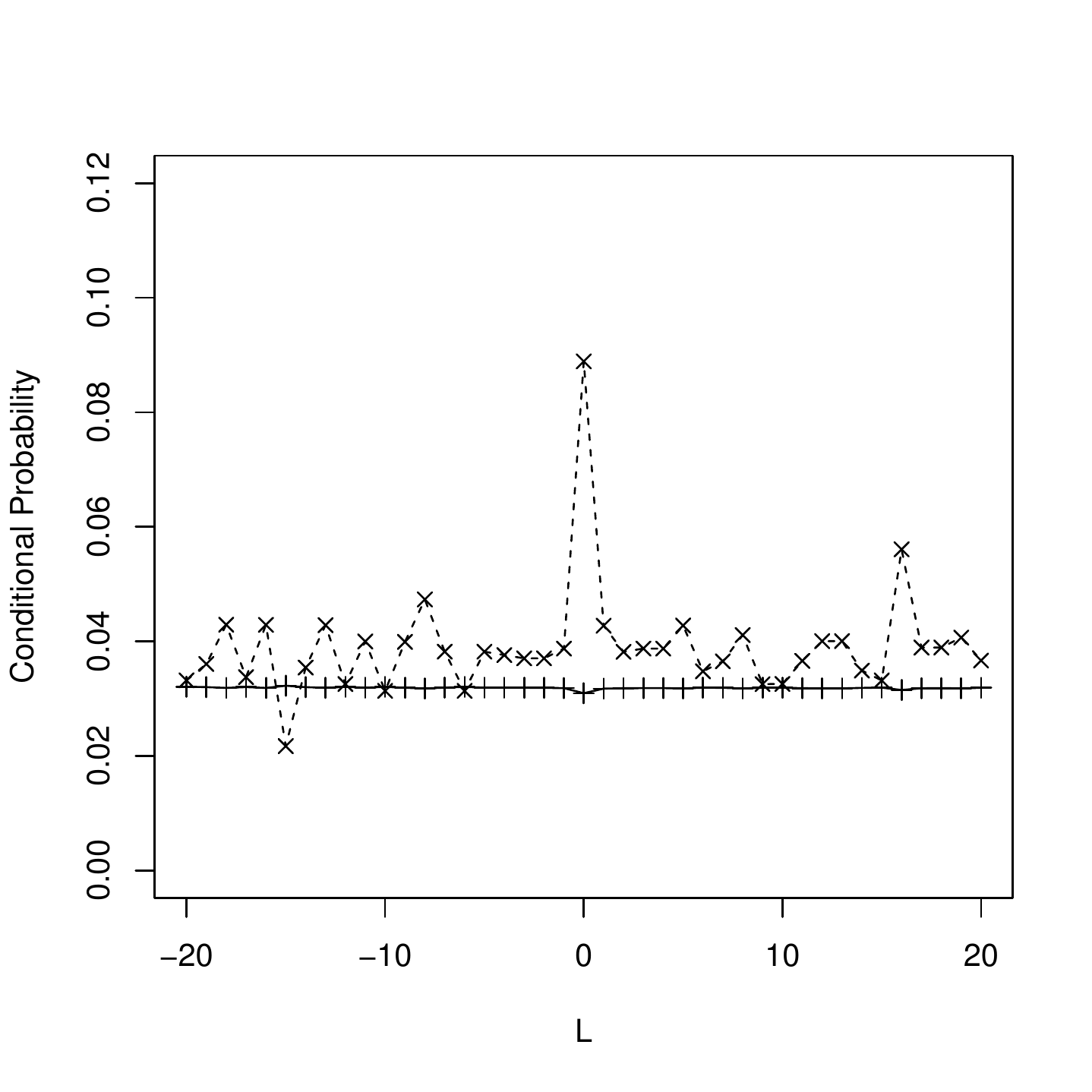}}
 
{\bf{Case 1-- Rewarded}}

 \end{center}  
  \begin{center}
  \subfigure[]{\includegraphics[width=2.7in,height=1.4in]{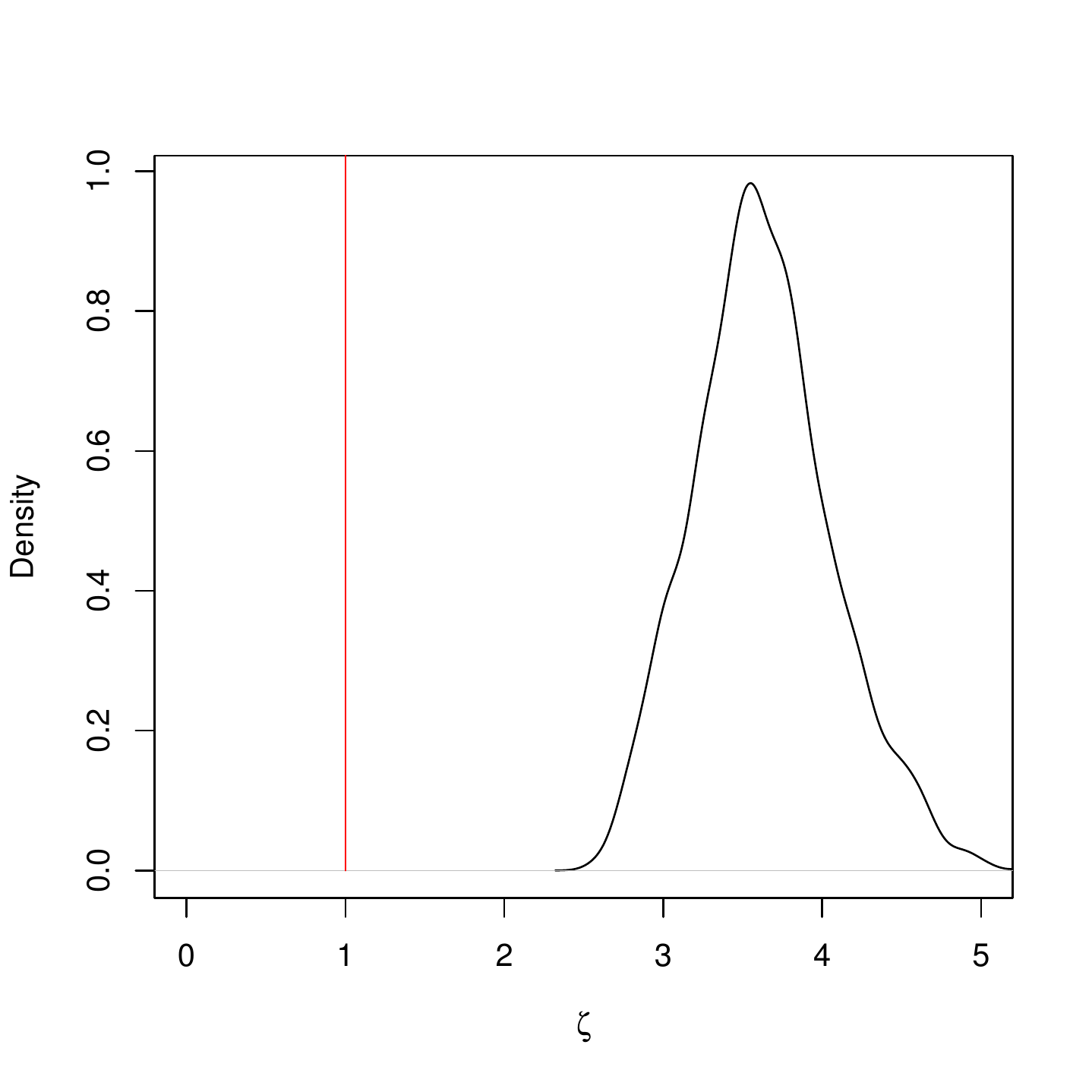}}\label{1a}
  \subfigure[]{\includegraphics[width=2.7in,height=1.4in]{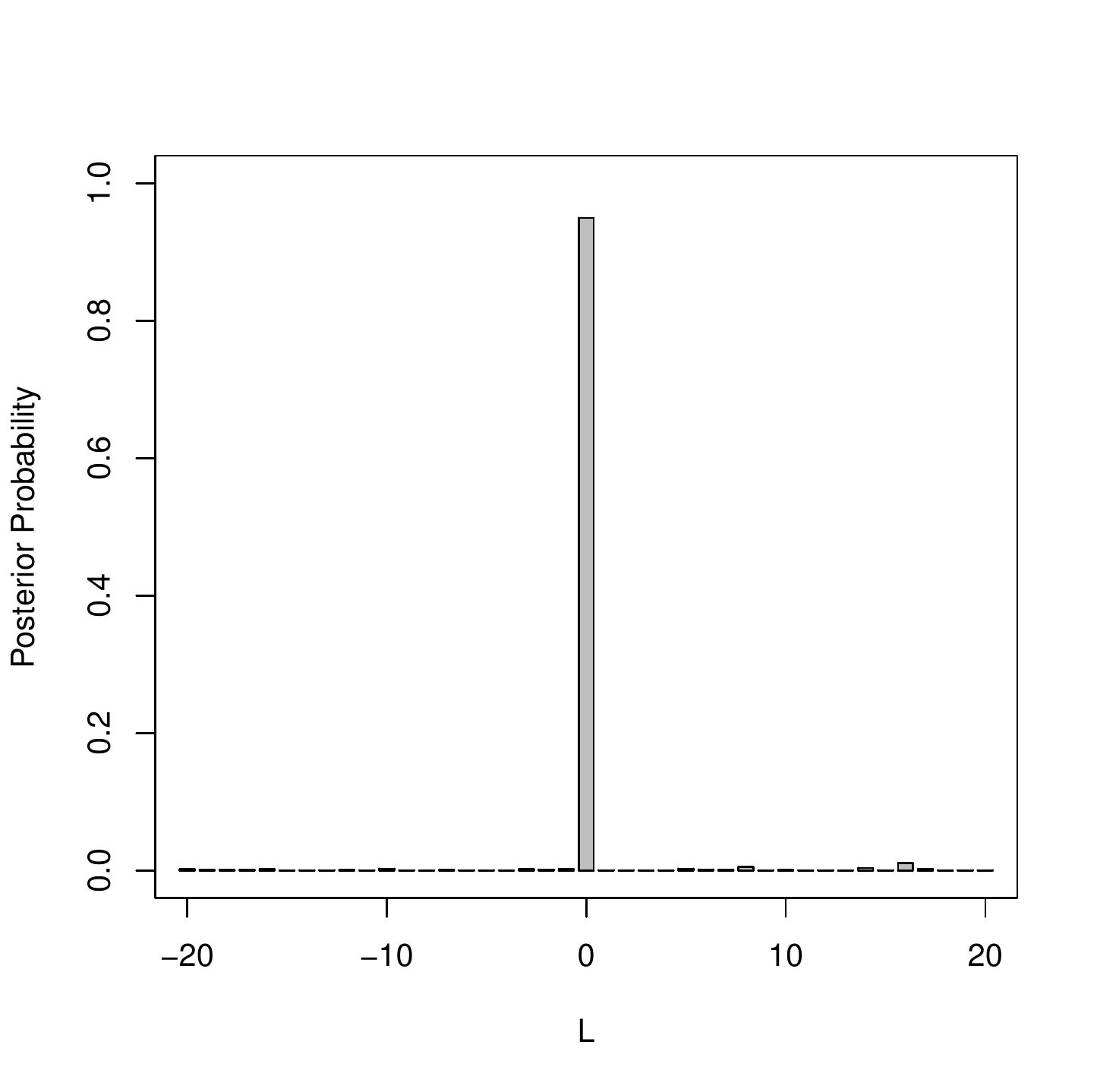}}\label{1b}

\subfigure[]{\includegraphics[width=1.8in, height=1.4in]{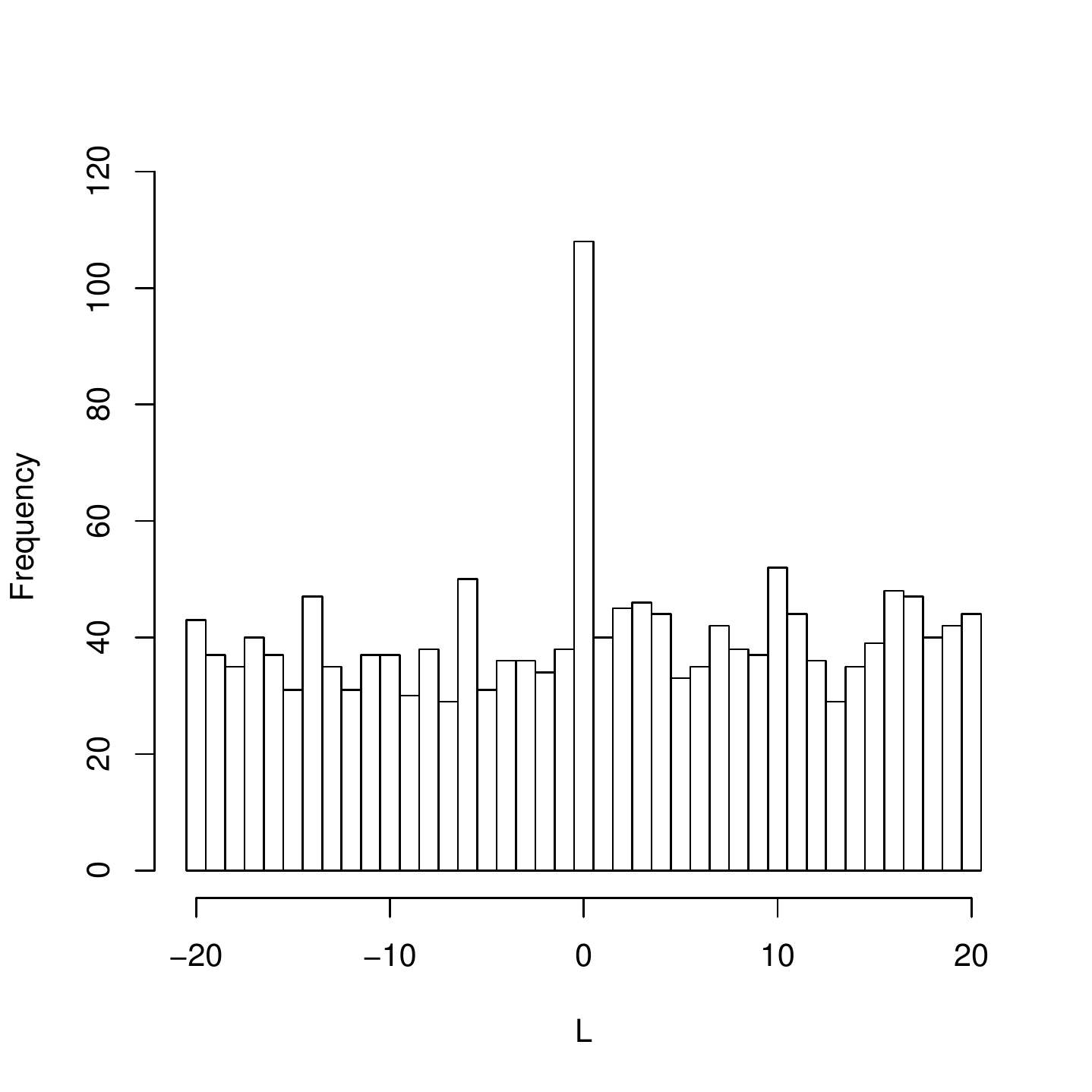}} \subfigure[]{\includegraphics[width=1.8in, height=1.4in]{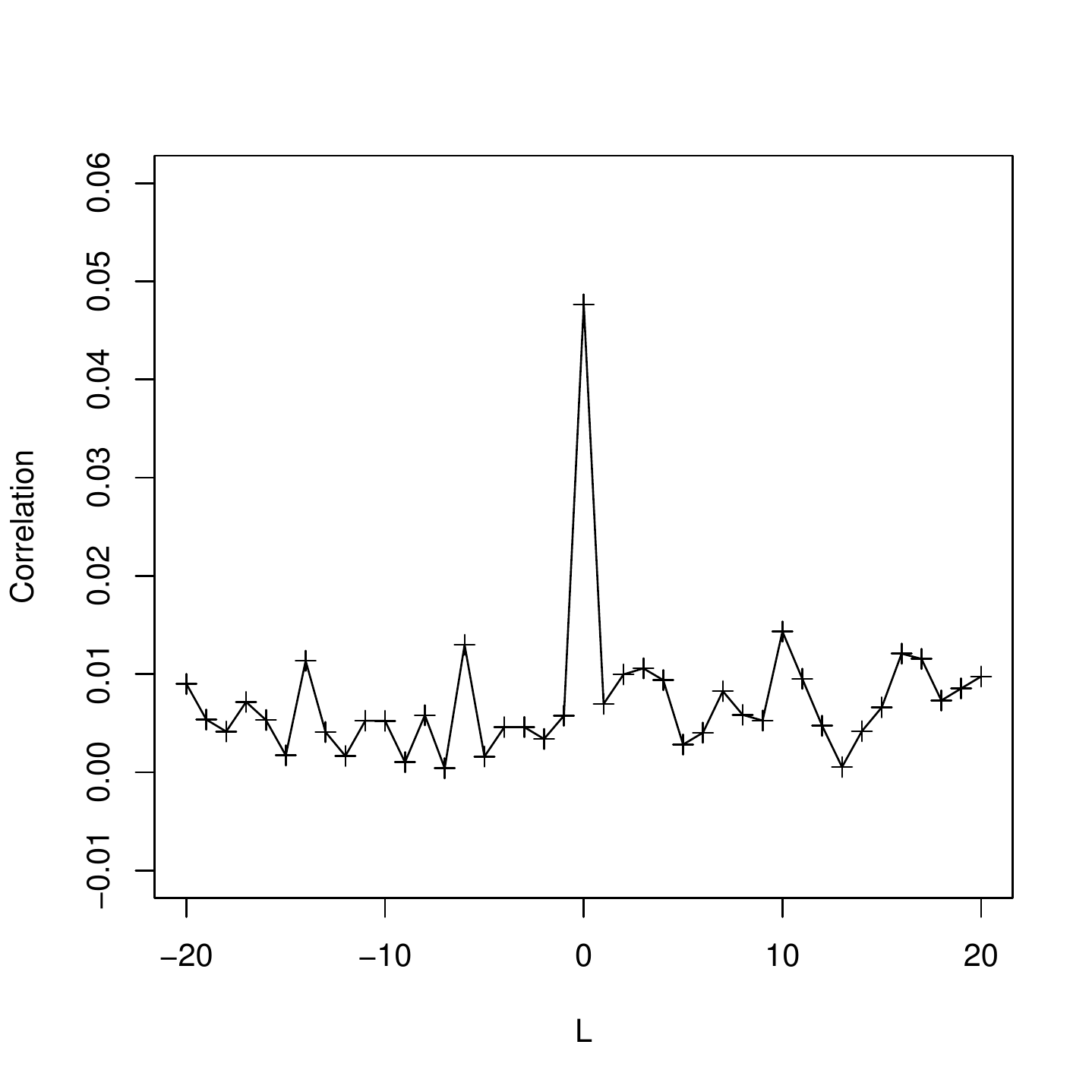}} \subfigure[]{\includegraphics[width=1.8in, height=1.4in]{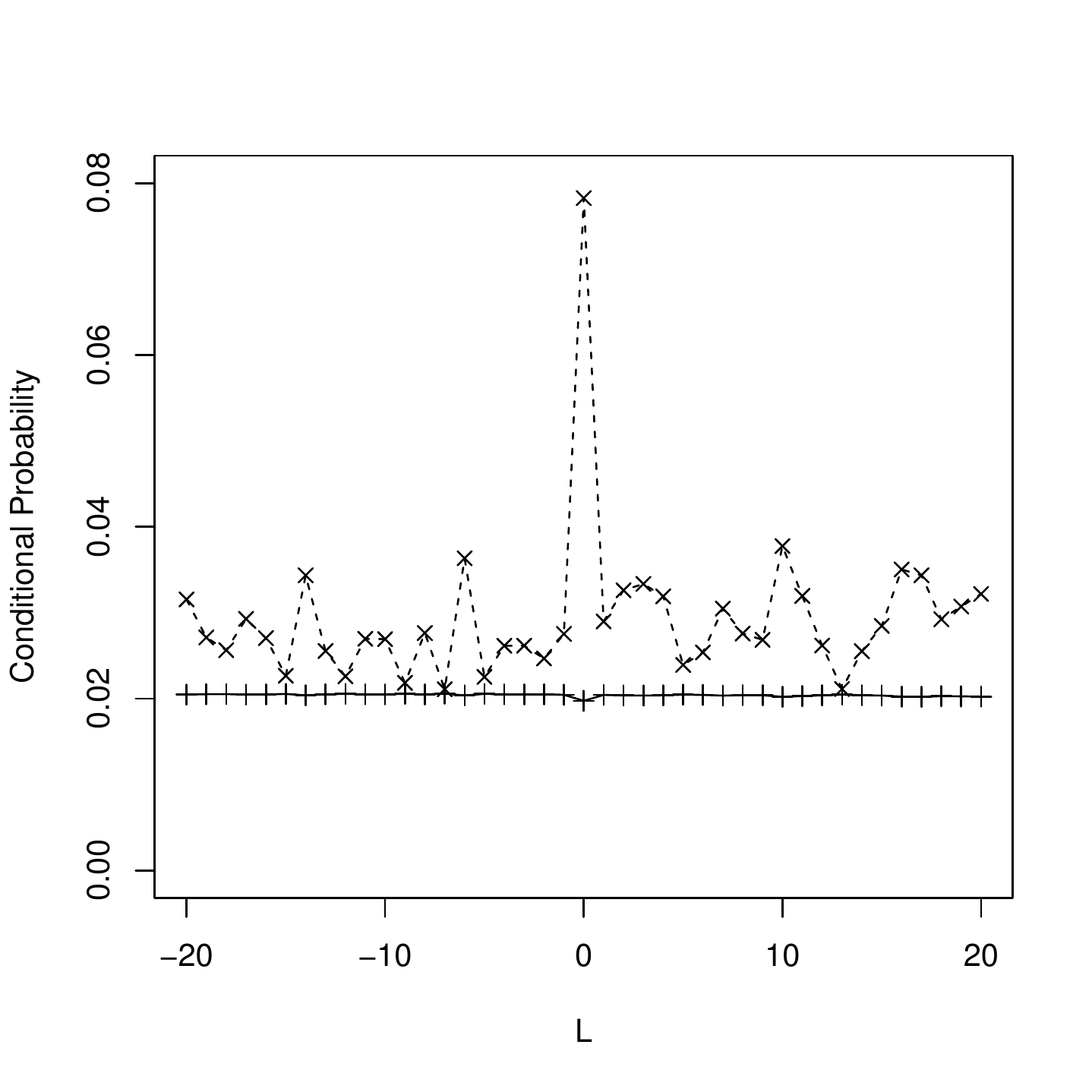}}
 {\bf{Case 1-- Non-rewarded}}
 \caption{{\bf Case 1}: a) Posterior distribution of $\zeta$, b) posterior distribution of lag, c) co-firing frequencies, d) correlation coefficients, and e) estimated conditional probabilities of firing for the second neuron given the firing status (0: solid line, 1: dashed line) of the first neuron over different lag values for the rewarded scenario; (f)-(j) are the corresponding plots for the non-rewarded scenario.}\label{case1Data}
 \end{center}

\end{figure}

We first present our model's results for a pair of neurons appear to be in exact synchrony under both scenarios. Figure \ref{case1Data} shows the posterior distributions of $\zeta$ and $L$ under different scenarios. As we can see, the posterior distributions of $\zeta$ in both cases are away from 1, and $L=0$ has the highest posterior probability. These results are further confirmed by empirical results, namely, the number of co-firings, correlation coefficients, and the sample estimates of conditional probabilities presented in Figure \ref{case1Data}. 

Using the method of \cite{kass11}, the p-values under the two scenarios are $3.2E-11$ and $1.4E-13$ respectively. While both methods provide similar conclusions, their method is designed to detect exact synchrony only.

\begin{figure}[!t]
  \begin{center}
  \subfigure[]{\includegraphics[width=2.7in,height=1.4in]{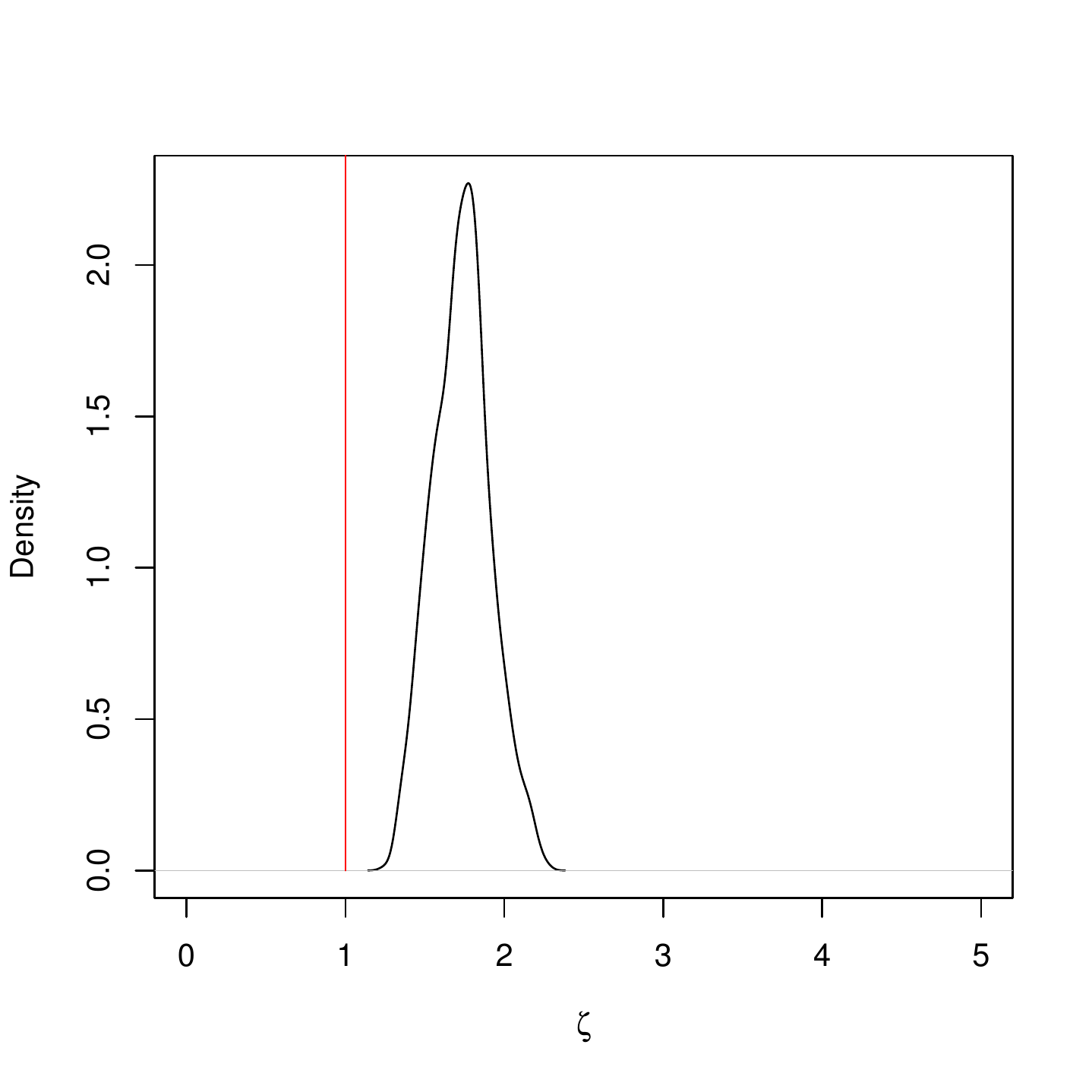}}
  \subfigure[]{\includegraphics[width=2.7in,height=1.4in]{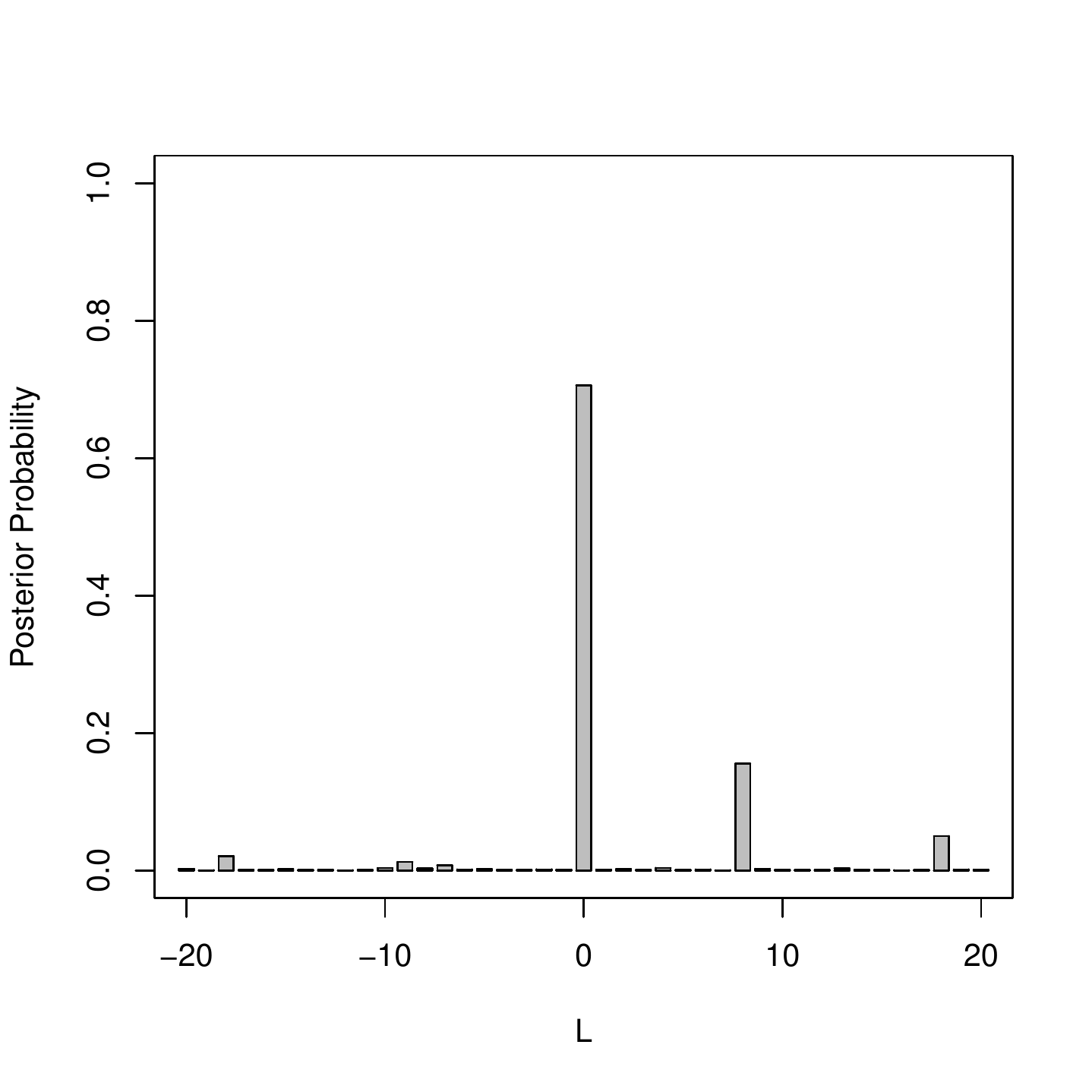}}

\subfigure[]{\includegraphics[width=1.8in, height=1.4in]{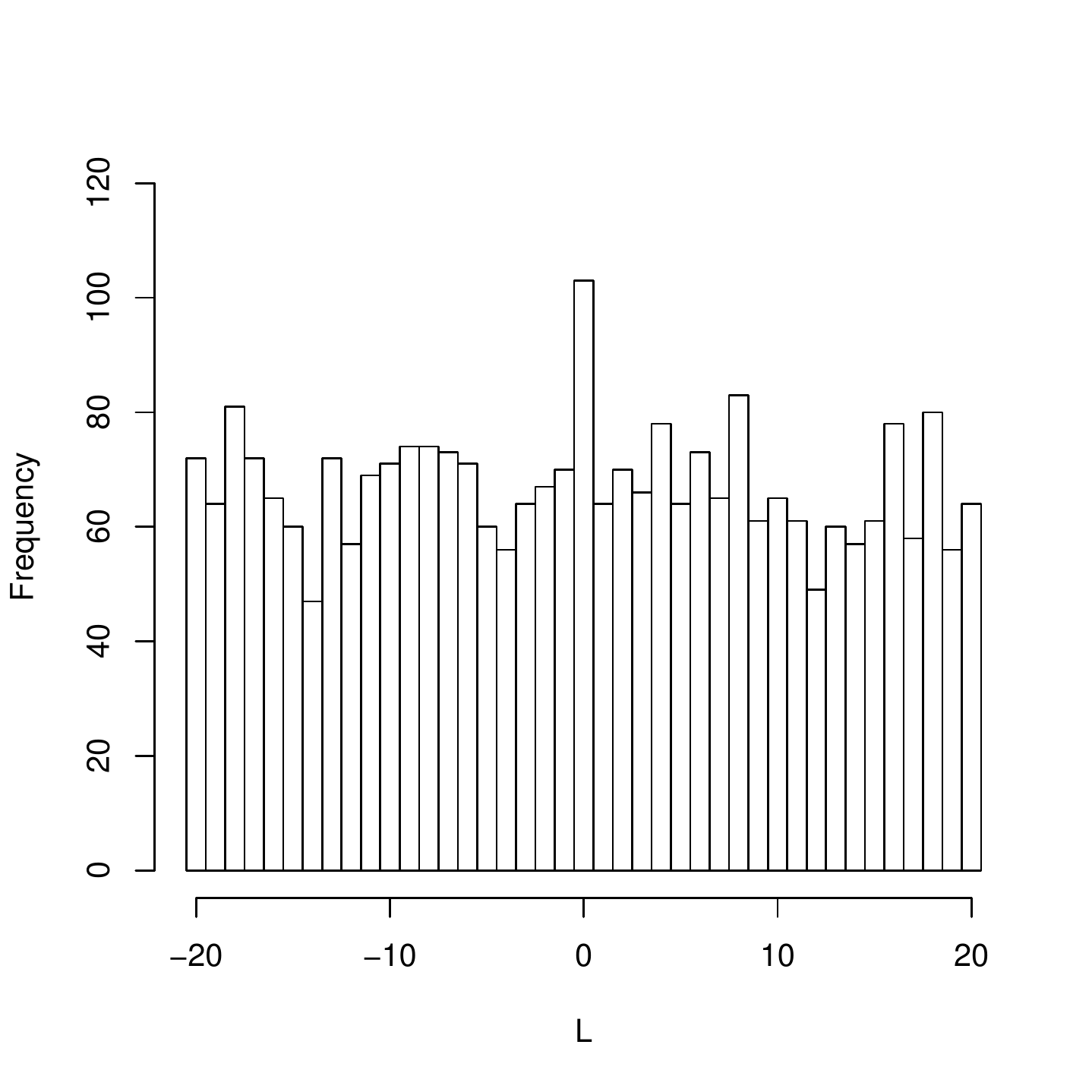}} \subfigure[]{\includegraphics[width=1.8in, height=1.4in]{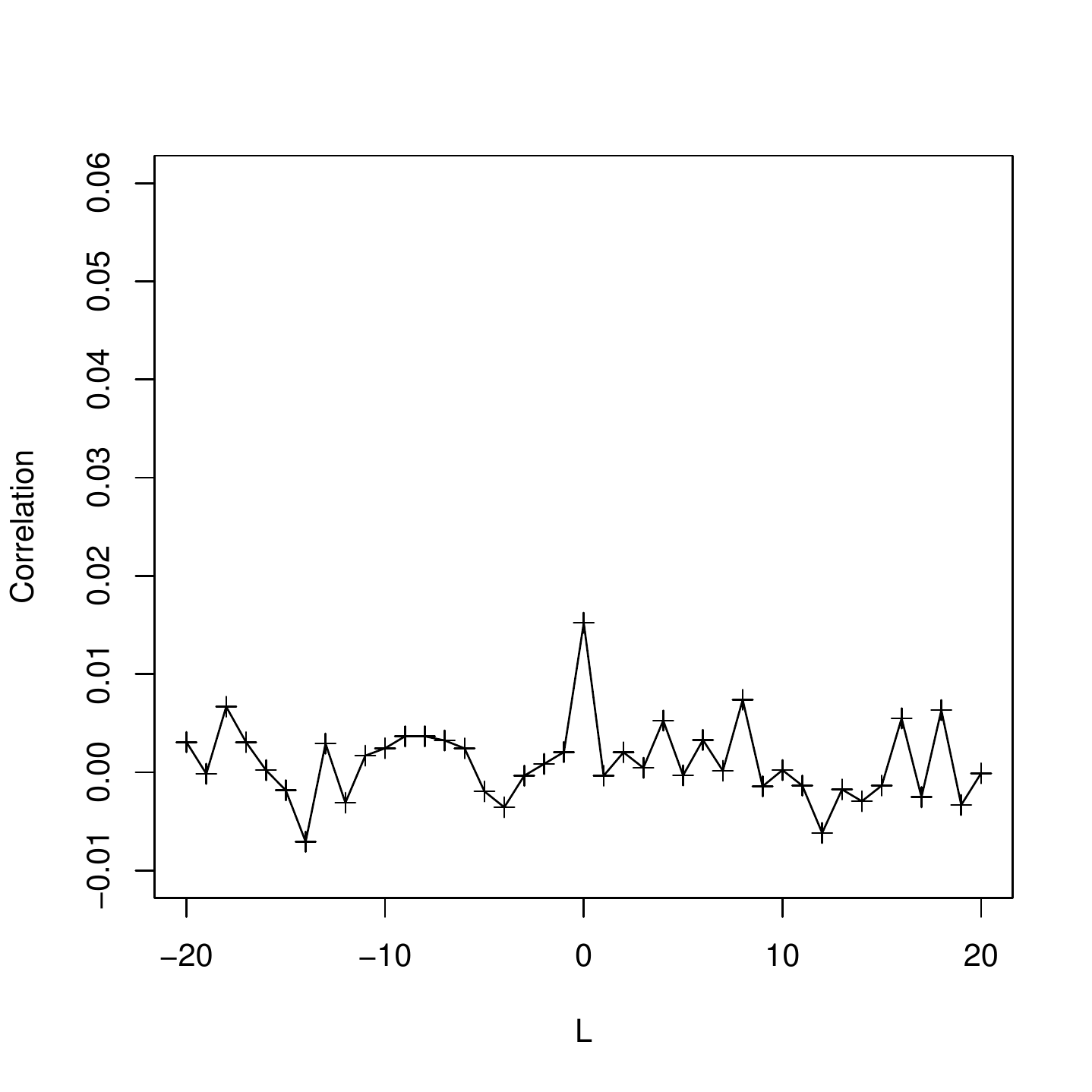}} \subfigure[]{\includegraphics[width=1.8in, height=1.4in]{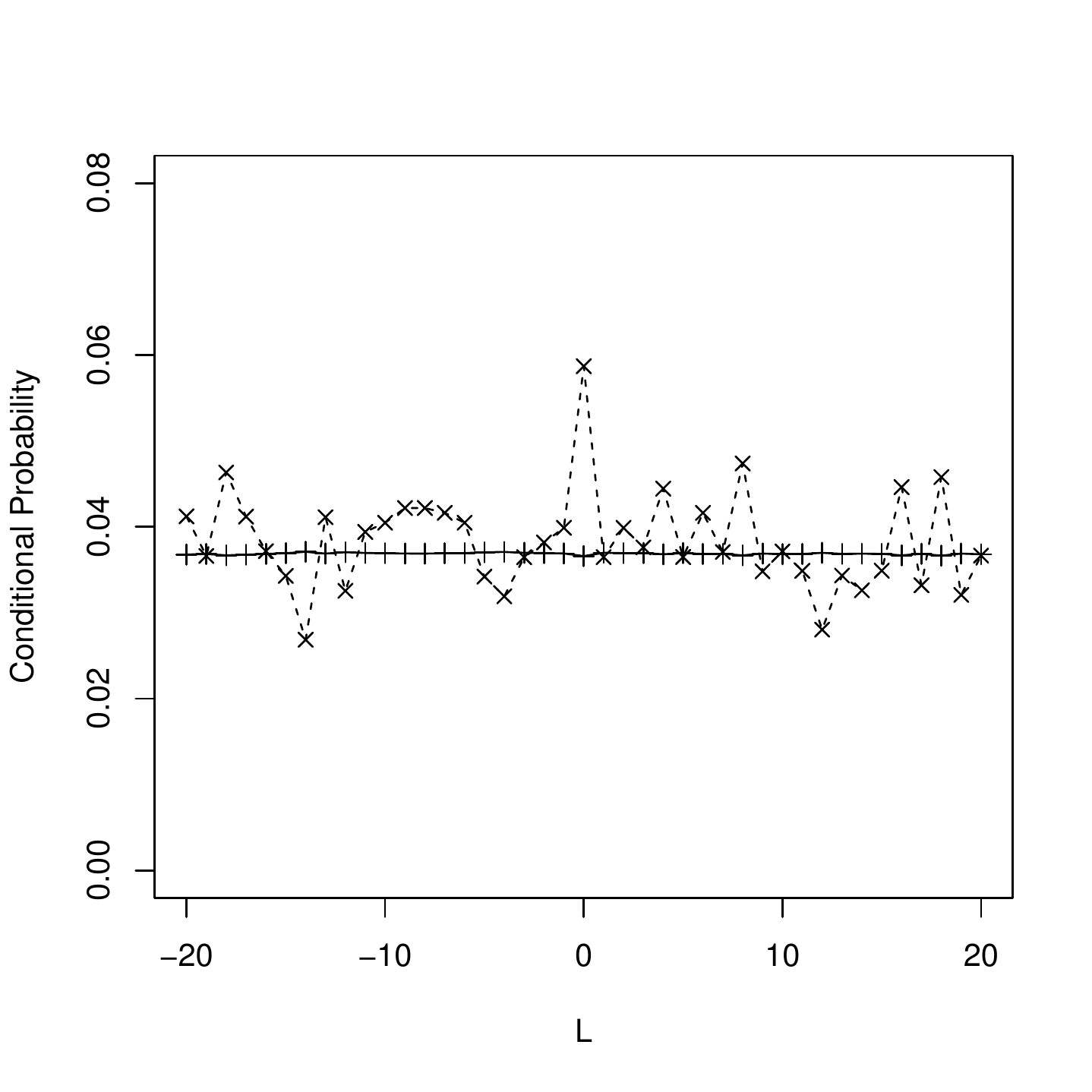}}
 
{\bf{Case 2-- Rewarded}}

 \end{center}  

  \begin{center}
  \subfigure[]{\includegraphics[width=2.7in,height=1.4in]{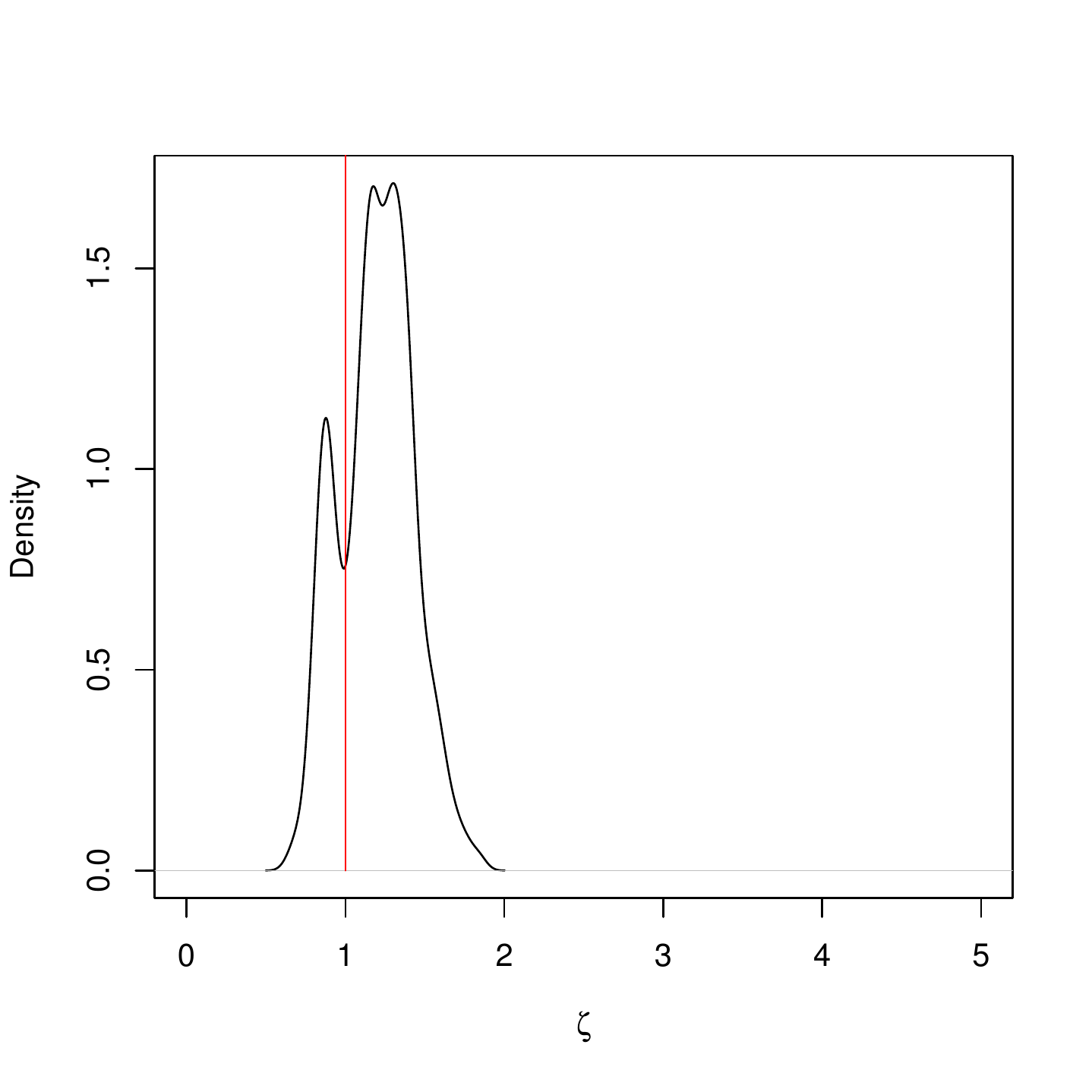}}
  \subfigure[]{\includegraphics[width=2.7in,height=1.4in]{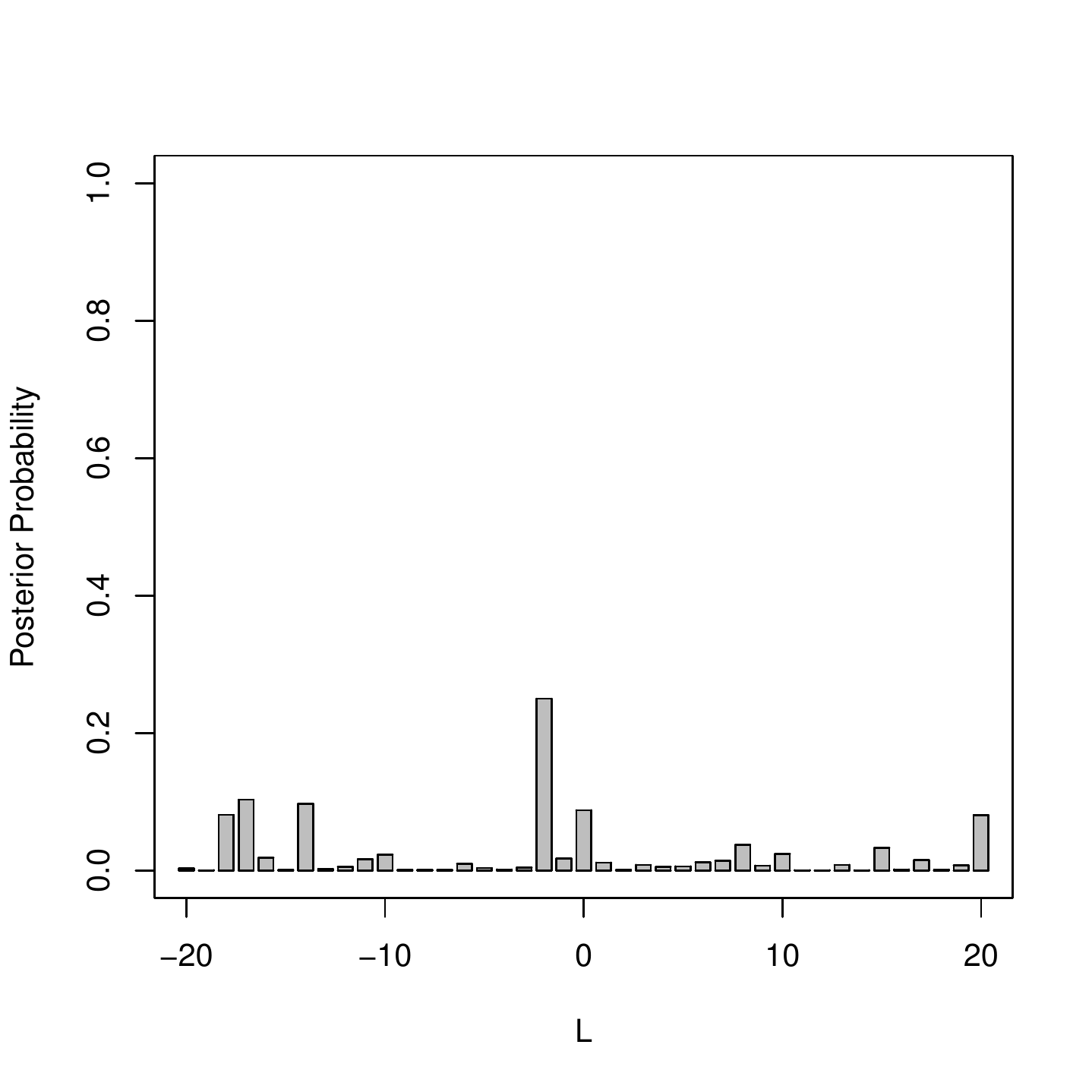}}

\subfigure[]{\includegraphics[width=1.8in, height=1.4in]{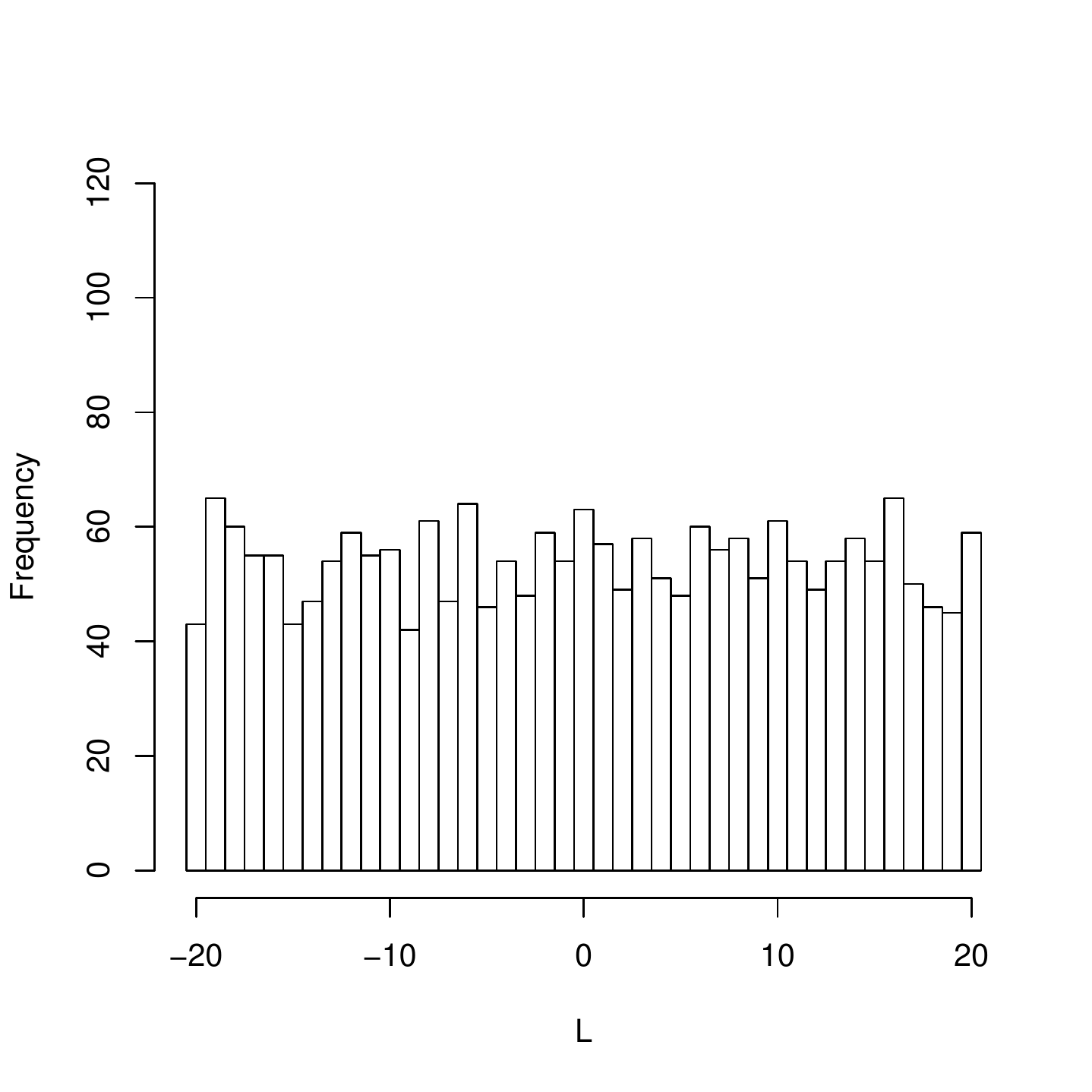}} \subfigure[]{\includegraphics[width=1.8in, height=1.4in]{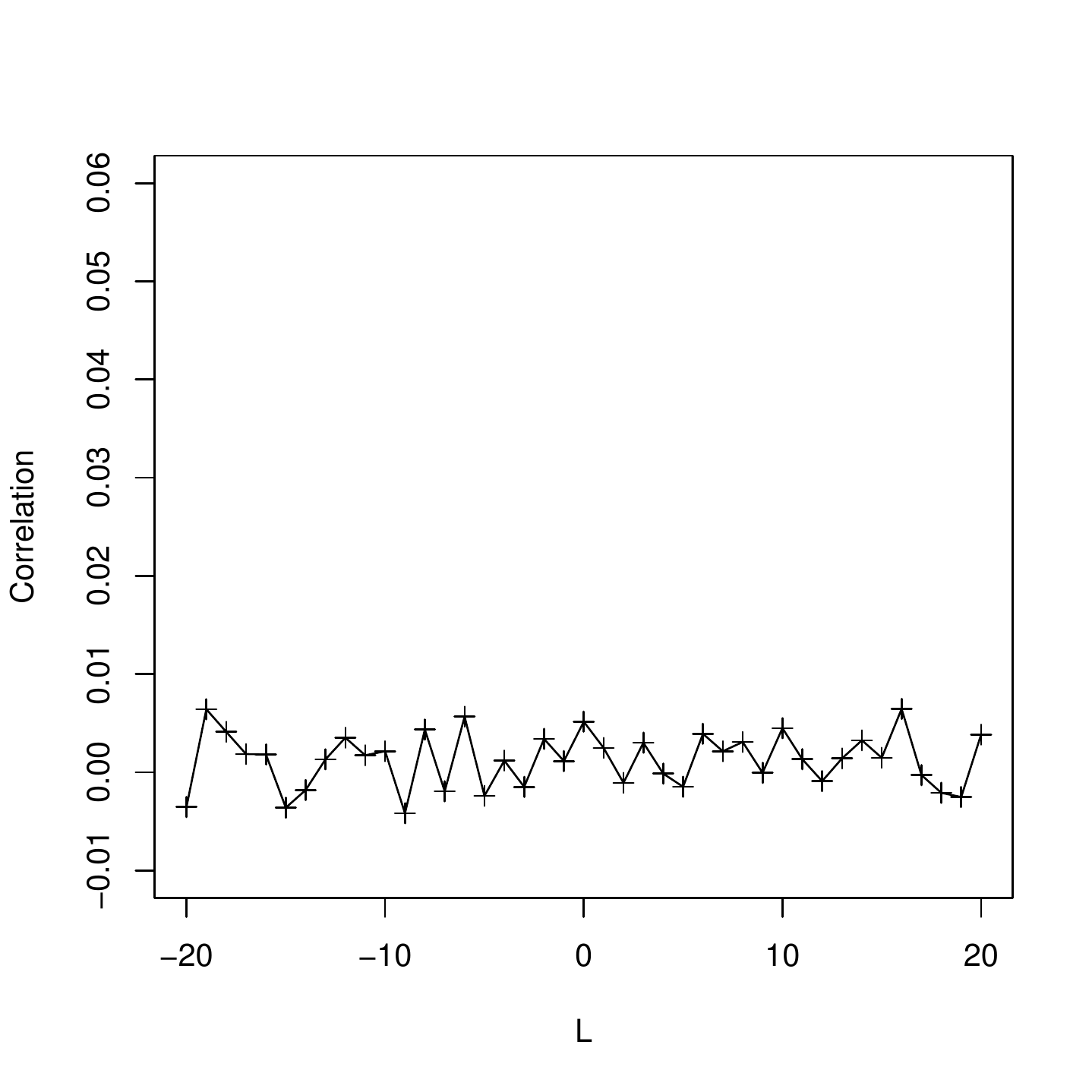}} \subfigure[]{\includegraphics[width=1.8in, height=1.4in]{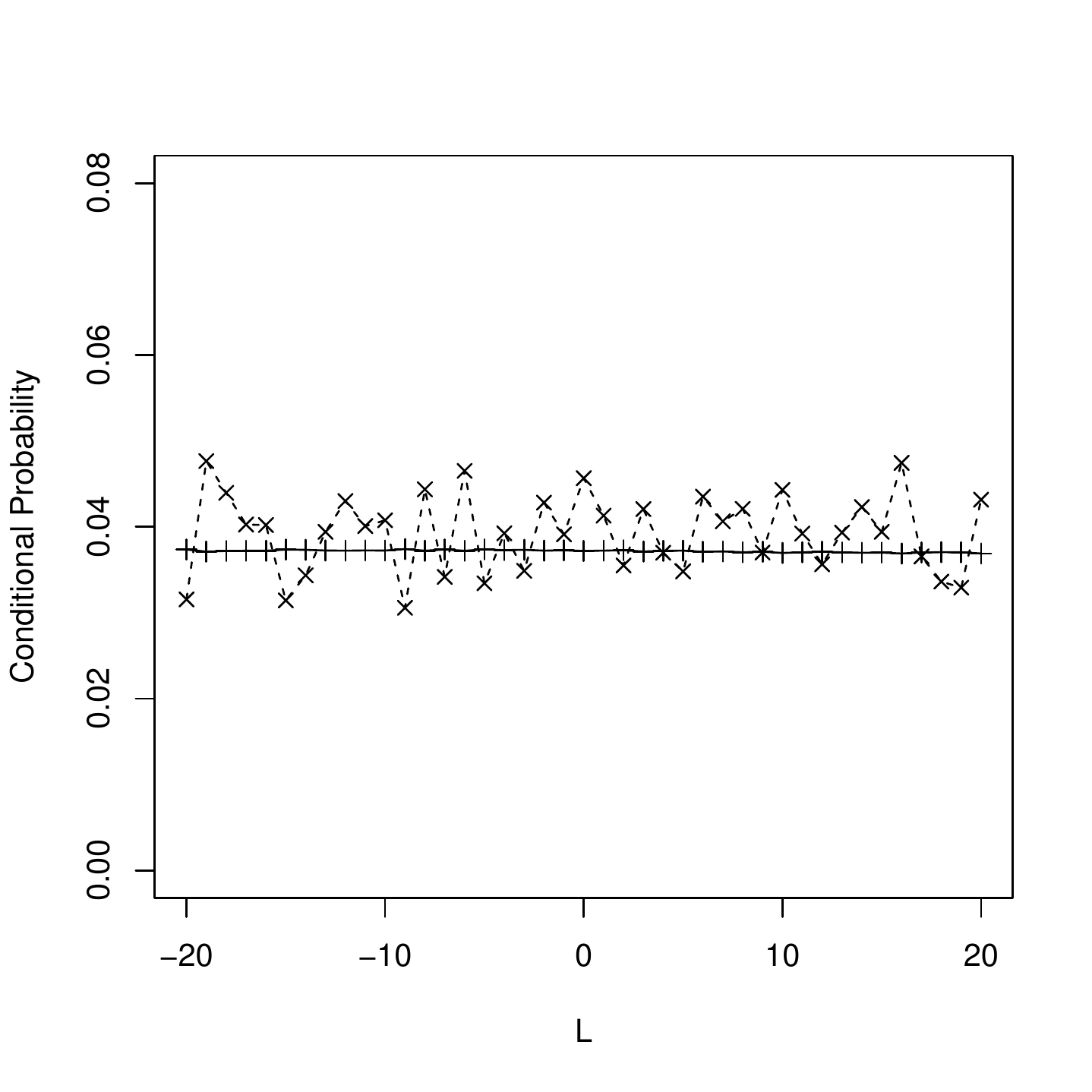}}
 {\bf{Case 2-- Non-rewarded}}
 \caption{{\bf Case 2}: a) Posterior distribution of $\zeta$, b) posterior distribution of lag, c) co-firing frequencies, d) correlation coefficients, and e) estimated conditional probabilities of firing for the second neuron given the firing status (0: solid line, 1: dashed line) of the first neuron over different lag values for the rewarded scenario; (f)-(j) are the corresponding plots for the non-rewarded scenario.}\label{case2Data}
 \end{center}  
\end{figure}

\subparagraph{Case 2: Two neurons with synchrony under the rewarded scenario only}

Next, we present our model's results for a pair of neurons appear to be in a moderate synchrony under the rewarded scenario only. Figure \ref{case2Data} shows the posterior distributions of $\zeta$ and $L$ under different scenarios. In this case, the posterior distributions of $\zeta$ in slightly away from 1 in the first scenario; however, under the second scenario, the tail probability of 1 is not negligible. These results are further confirmed by empirical results presented in Figure \ref{case2Data}: only in the first scenario we observe a moderate difference between the conditional probabilities. 

Using the method of \cite{kass11}, the p-values under the two scenarios are $2E-4$ and $0.144$ respectively. As discussed above, although for these data the two methods provide similar results in terms of synchrony, our
method can be used to make inference regarding possible lag values. Moreover, as we will show in the next section, our method provides a hierarchical Bayesian framework that can be easily extended to multiple neurons.

\section{Modeling dependencies among multiple neurons}\label{multipleNeurons}
Temporal relationships among neurons, particularly those that change across different contexts, can provide additional information beyond basic firing rate. Because it is possible to record spike trains from multiple neurons simultaneously, and because network encoding likely spans more than pairs of neurons, we now turn our attention to calculating temporally-related activity among multiple ($>2$) simultaneously-recorded neurons.

At lag zero (i.e., $L=0$), we can rewrite our model for the joint distribution of two neurons in terms of their individual cumulative distributions as follows (we have dropped the index $t$ for simplicity):
\begin{eqnarray*}
H(y, z)= \mathcal{C}(F_1(y),F_2(z))=\big[1+\beta\prod_{i=1}^2(1-F_i) \big]\prod_{i=1}^2F_i
\end{eqnarray*}
where $F_1=F_1(y)=P(Y\le y)$, $F_2=F_2(z)=P(Z\le z)$, and $\beta=\frac{\zeta-1}{(1-p)(1-q)}$. Note that in this case, $\beta=0$ indicates that the two neurons are independent. In general, models that couple the joint distribution of two (or more) variables to their individual marginal distributions are called copula models. See \cite{nelsen98} for detailed discussion of copula models. Let $H$ be $n$-dimensional distribution functions with marginals $F_1,...,F_n$. Then, an $n$-dimensional copula is a function of the following form:
\begin{eqnarray*}
H(y_1,...,y_n) = \mathcal{C}(F_1(y_1),...,F_n(y_n)), \ \textrm{for all } y_1, \ldots,y_n
\end{eqnarray*}
Here, $\mathcal{C}$ defines the dependence structure between the marginals. Our model for two neurons is in fact a special case of the Farlie-Gumbel-Morgenstern (FGM) copula family \citep{farlie60, gumbel60, morgenstern56, nelsen98}. For $n$ random variables $Y_{1}, Y_{2}, \ldots, Y_{n}$, the FGM copula, $\mathcal{C}$, has the following form:
\begin{eqnarray}
\mathcal{C} = \big[1+\sum_{k=2}^n \ \ \sum_{1\le j_1< \cdots <j_k\le n}\beta_{j_1j_2...j_k}\prod_{l=1}^k(1-F_{j_l}) \big] \prod_{i=1}^nF_i
\end{eqnarray}
where $F_i=F_i(y_i)$. Restricting our model to second-order interactions, we can generalize our approach for two neurons to a copula-based model for multiple neurons using the FGM copula family,
\begin{eqnarray}\label{copula}
H(y_1, \ldots ,y_n)=\big[1+\sum_{1\le j_1<j_2\le n}\beta_{j_1j_2}\prod_{l=1}^2(1-F_{j_l})\big] \prod_{i=1}^nF_i
\end{eqnarray}
where $F_i=P(Y_i\le y_i)$. Here, we use $y_{1}, \ldots, y_{n}$ to denote the firing status of $n$ neurons at time $t$; $\beta_{j_1 j_2}$ captures the relationship between the $j_1^{th}$ and $j_2^{th}$ neurons. To ensure that probability distribution functions remain within $[0, 1]$, the following constraints on all $n\choose 2$ parameters $\beta_{j_1j_2}$ are imposed:
\begin{equation*}
1+\sum_{1\leq j_1<j_2\leq n} \beta_{j_1j_2}\prod_{l=1}^2 \epsilon_{j_l}\geq 0, \quad\epsilon_1,\cdots, \epsilon_n\in\{-1,1\}
\end{equation*}
Considering all possible combinations of $\epsilon_{j_1}$ and
$\epsilon_{j_2}$ in the above condition, there are $n(n-1)$ linear inequalities, which can be combined into the following inequality:
\begin{equation*}
\sum_{1\leq j_1<j_2\leq n} |\beta_{j_1j_2}| \leq 1
\end{equation*}

\begin{table}[t]%\scriptsize
\caption{{Estimates of $\beta$'s along with their 95\% posterior probability intervals for simulated data based on our copula-based model. Here, row $i$ column $j$ shows $\beta_{ij}$, which captures the relationship between the $i^{th}$ and $j^{th}$ neurons.}}
\label{EffectsResultsSim}
\begin{center}
%\begin{scriptsize}
\begin{tabular}{|c||c|c|}
\hline
$\beta$ & 2 & 3 \\
\hline
\hline
1  & {\bf{0.66 (0.30,0.94)}} & 0.02 (-0.26,0.27) \\ \hline
2                   & & -0.05 (-0.33,0.19)\\ \hline
\end{tabular}
%\end{scriptsize}
\end{center}
\end{table}

\subsection{Illustrative example}
To illustrate this method, we follow a similar procedure as Section \ref{sec:illust} and simulate spike trains for three neurons such that neurons 1 and 2 are in exact synchrony, but they are independent from neuron 3. 
Table \ref{EffectsResultsSim} shows the estimated $\beta$'s along with their corresponding 95\% posterior probability intervals using posterior samples from Spherical HMC. Our method correctly detects the relationship among the neurons: for synchronous neurons, the corresponding $\beta$'s are significantly larger than 0 (i.e., 95\% posterior probability intervals do not include 0), whereas the remaining $\beta$'s are close to 0 (i.e., 95\% posterior probability intervals include 0).

\subsection{Results for experimental data}
We now use our copula-based method for analyzing the experimental data discussed earlier. As mentioned, during task performance the activity of multiple neurons was recorded under two conditions: rewarded stimulus (lever 1) and non-rewarded stimulus (lever 2). Here, we focus on 5 simultaneously recorded neurons. There are 51 trials per neuron under each scenario. We set the time intervals to 5 ms. 

\begin{figure}[t]
  \begin{center}
  \includegraphics[width=2.5in,height=2in]{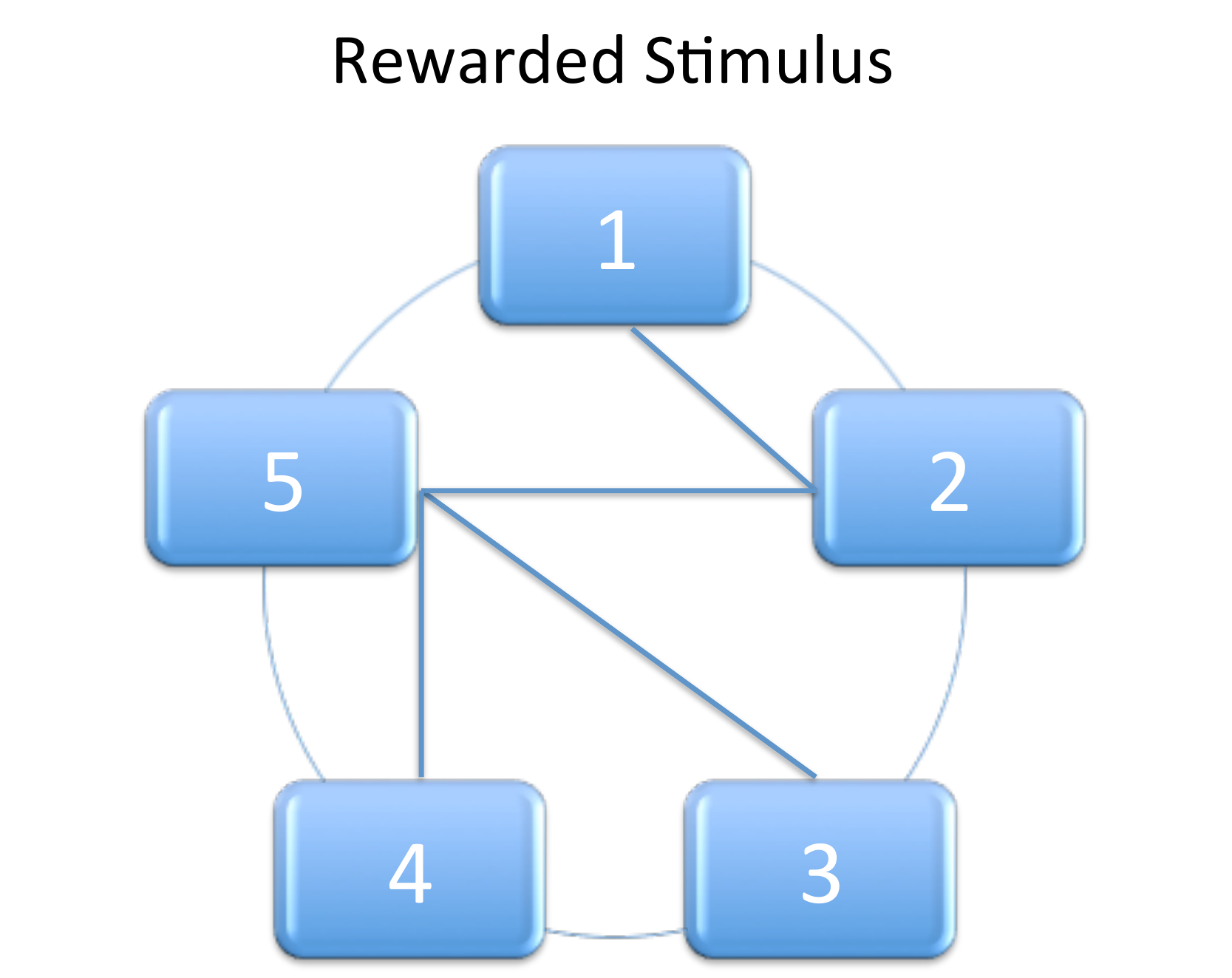} \hspace{20pt} \includegraphics[width=2.5in,height=2in]{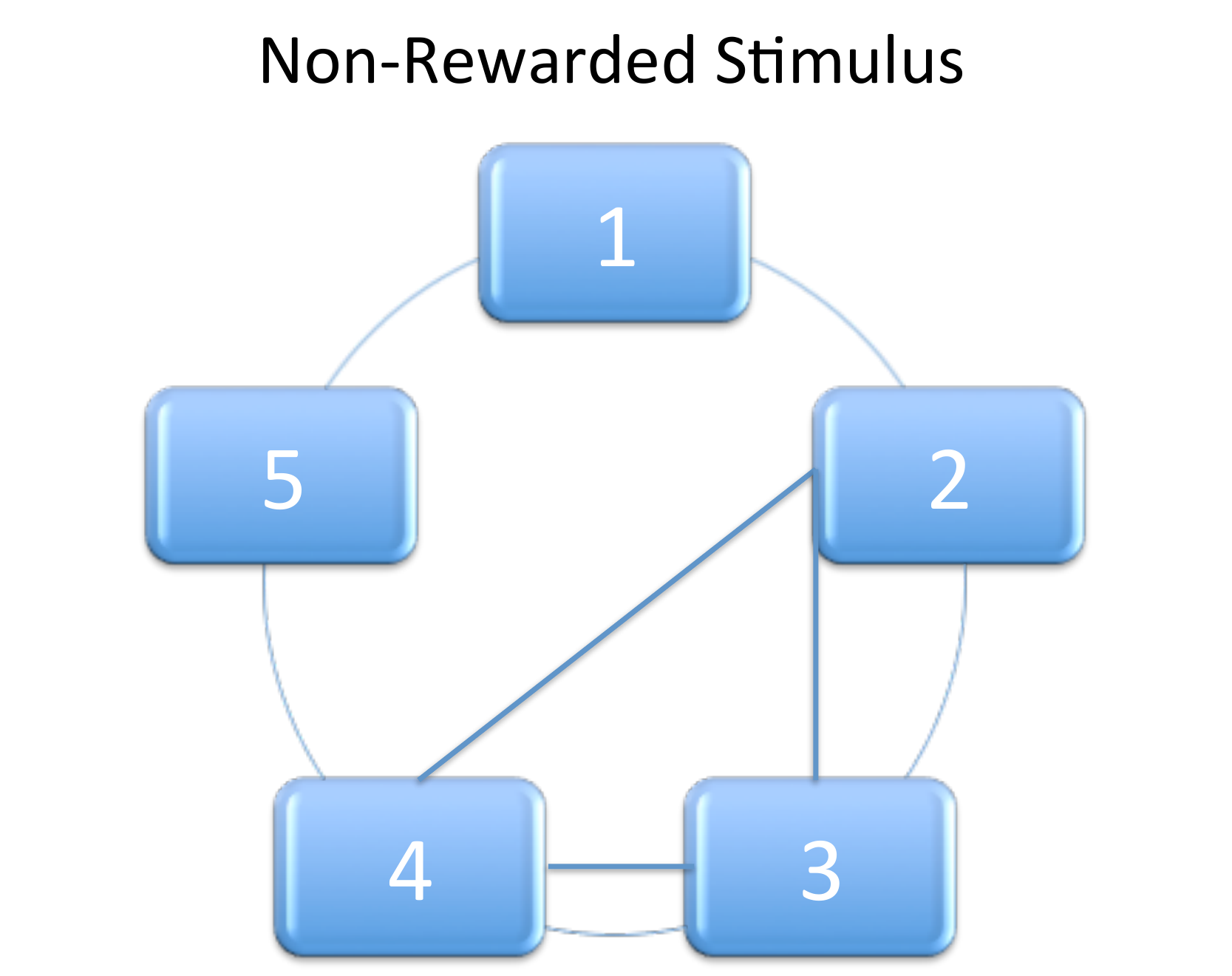}
  \caption{A schematic representation of connections between five neurons under two experimental conditions. The solid line indicates significant association. }\label{schematic}
  \end{center}
\end{figure}

Tables \ref{EffectsResults1} and \ref{EffectsResults2} show the estimates of $\beta_{i,j}$, which capture the association between the $i^{th}$ and $j^th$ neurons, under the two scenarios. Figure \ref{schematic} shows the schematic representation of these results under the two experimental conditions. The solid line indicates significant association.

\begin{table}[t]
\begin{center}
\caption{{Estimates of $\beta$'s along with their 95\% posterior probability intervals for the first scenario (Rewarded) based on our copula model.}}
\label{EffectsResults1}
\vspace{6pt}
\begin{tabular}{|c||c|c|c|c|}
\hline
$\beta$ & 2 & 3 & 4 & 5\\
\hline
\hline
1  & {\bf{0.22(0.07,0.39)}} & 0.00(-0.07,0.04) & 0.03(-0.02,0.15)& 0.01(-0.04,0.08) \\ \hline
2  & & 0.03(-0.02,0.18) & 0.06(-0.02,0.22) & {\bf{0.07(0.00,0.25)}}   \\ \hline
3 & & & 0.08(-0.01,0.26)& {\bf{0.21(0.04,0.38)}}  \\ \hline
4 & & & &{\bf{0.23(0.09,0.40)}}  \\ \hline
\end{tabular}
\end{center}
\begin{center}
\caption{{Estimates of $\beta$'s along with their 95\% posterior probability intervals for the second scenario (Non-rewarded) based on our copula model.}}
\label{EffectsResults2}
\vspace{6pt}
\begin{tabular}{|c||c|c|c|c|}
\hline
$\beta$ & 2 & 3 & 4 & 5\\
\hline
\hline
1 & 0.05(-0.02,0.25) & -0.01(-0.09,0.04) & 0.15(-0.01,0.37)& 0.05(-0.03,0.22) \\ \hline
2 & & {\bf{0.21(0.03,0.41)}} & {\bf{ 0.18(0.00,0.37)}} & 0.03(-0.02,0.19)  \\ \hline
3 &  & & {\bf{0.17(0.00,0.34)}} & 0.03(-0.02,0.19) \\ \hline
4 & & & & 0.07(-0.01,0.24)  \\ \hline
\end{tabular}
\end{center}
\end{table}

Our results show that neurons recorded simultaneously in the same brain area are correlated in some conditions and not others. This strongly supports the hypothesis that population coding among neurons (here though correlated activity) is a meaningful way of signaling differences in the environment (rewarded or non-rewarded stimulus) or behavior (going to press the rewarded lever or not pressing) \cite{buzsaki10}. It also shows that neurons in the same brain region are differentially involved in different tasks, an intuitive perspective but one that is neglected by much of behavioral neuroscience. Finally, our results indicate that network correlation is dynamic and that functional pairs-- again, even within the same brain area-- can appear and disappear depending on the environment or behavior. This suggests (but does not confirm) that correlated activity across separate populations within a single brain region can encode multiple aspects of the task. For example, the pairs that are correlated in reward and not in non-reward could be related to reward-seeking whereas pairs that are correlated in non-reward could be related to response inhibition.  Characterizing neural populations within a single brain region based on task-dependent differences in correlated firing is a less-frequently studied phenomenon compared to the frequently pursued goal of identifying the overall function of the brain region based on individual neural firing \cite{stokes13}. While our data only begin to address this important question, the developed model will be critical in application to larger neural populations across multiple tasks in our future research.

\section{Computation}\label{computation}
We use Markov Chain Monte Carlo (MCMC) algorithms to sample from posterior distribution. Algorithm \ref{mainAlg} in   Appendix shows the overall sampling procedure. We use the slice sampler \citep{neal03} for the hyperparameters controlling the covariance function of the Gaussian process model. More specifically, we use the ``stepping out'' procedure to find an interval around the current state, and then the ``shrinkage'' procedure to sample from this interval. For latent variables with Gaussian process priors, we use the elliptical slice sampling algorithm proposed by \cite{murray10}. The details are provided in Algorithm \ref{ellipSlice} in the appendix. 

Sampling from the posterior distribution of $\beta$'s in the copula model is quite challenging. As the number of neurons increases, simulating samples from the posterior distribution these parameters becomes difficult because of the imposed constraints \citep{pneal08, sherlock09, pneal12, brubaker12, pakman13}. We have recently developed a new Markov Chain Monte Carlo algorithm for constrained target distributions \citep{lanICML14} based on Hamiltonian Monte Carlo (HMC) \citep{HMCb, neal11}. 

In many cases, bounded connected constrained $D$-dimensional parameter spaces can be bijectively mapped to the $D$-dimensional unit ball ${\bf B}_0^D(1):=\{\theta\in\mathbb R^D: \Vert \theta\Vert_2
=\sqrt{\sum_{i=1}^D \theta_i^2}\leq 1\}$, where $\theta$ are parameters. Therefore, our method first maps the $D$-dimensional constrained domain of parameters to the unit ball. We then augment the original $D$-dimensional parameter $\theta$ with an extra auxiliary variable $\theta_{D+1}$ to form an extended $(D+1)$-dimensional parameter $\tilde \theta = (\theta, \theta_{D+1})$ such that $\Vert \tilde\theta\Vert_2=1$ so $\theta_{D+1} = \pm \sqrt{1-\Vert \theta\Vert_2^2}$. This way, the domain of the target distribution is changed from the unit ball ${\bf B}_0^D(1)$ to the $D$-dimensional sphere, ${\bf S}^D :=\{\tilde \theta\in \mathbb R^{D+1}: \Vert \tilde\theta\Vert_2=1\}$, through the following transformation:
\begin{equation*}\label{b2s}
T_{{\bf B}\to {\bf S}}: {\bf B}_0^D(1)\longrightarrow {\bf S}^D, \; \theta \mapsto \tilde\theta = (\theta, \pm\sqrt{1-\Vert \theta\Vert_2^2})
\end{equation*}
Note that although $\theta_{D+1}$ can be either positive or negative, its sign does not affect our Monte Carlo estimates since after applying the above transformation, we need to adjust our estimates according to the change of variable theorem as follows:
\begin{equation*}\label{domains}
\int_{{\bf B}_0^D(1)} f(\theta) d\theta_{\bf B} = \int_{{\bf S}_+^D} f(\tilde\theta) \left|\frac{d\theta_{\bf B}}{d\tilde\theta_{\bf S}}\right| d\tilde\theta_{\bf S}
\end{equation*}
where $\left|\frac{d\theta_{\bf B}}{d\tilde\theta_{\bf S}}\right|=|\theta_{D+1}|$. Here, $d\theta_{\bf B}$ and $d\tilde\theta_{\bf S}$ are under Euclidean measure and spherical measure respectively. 

\begin{figure}[t]
\vspace{-10pt}
\begin{center}
\centerline{\includegraphics[width=3.2in, height=1.5in]{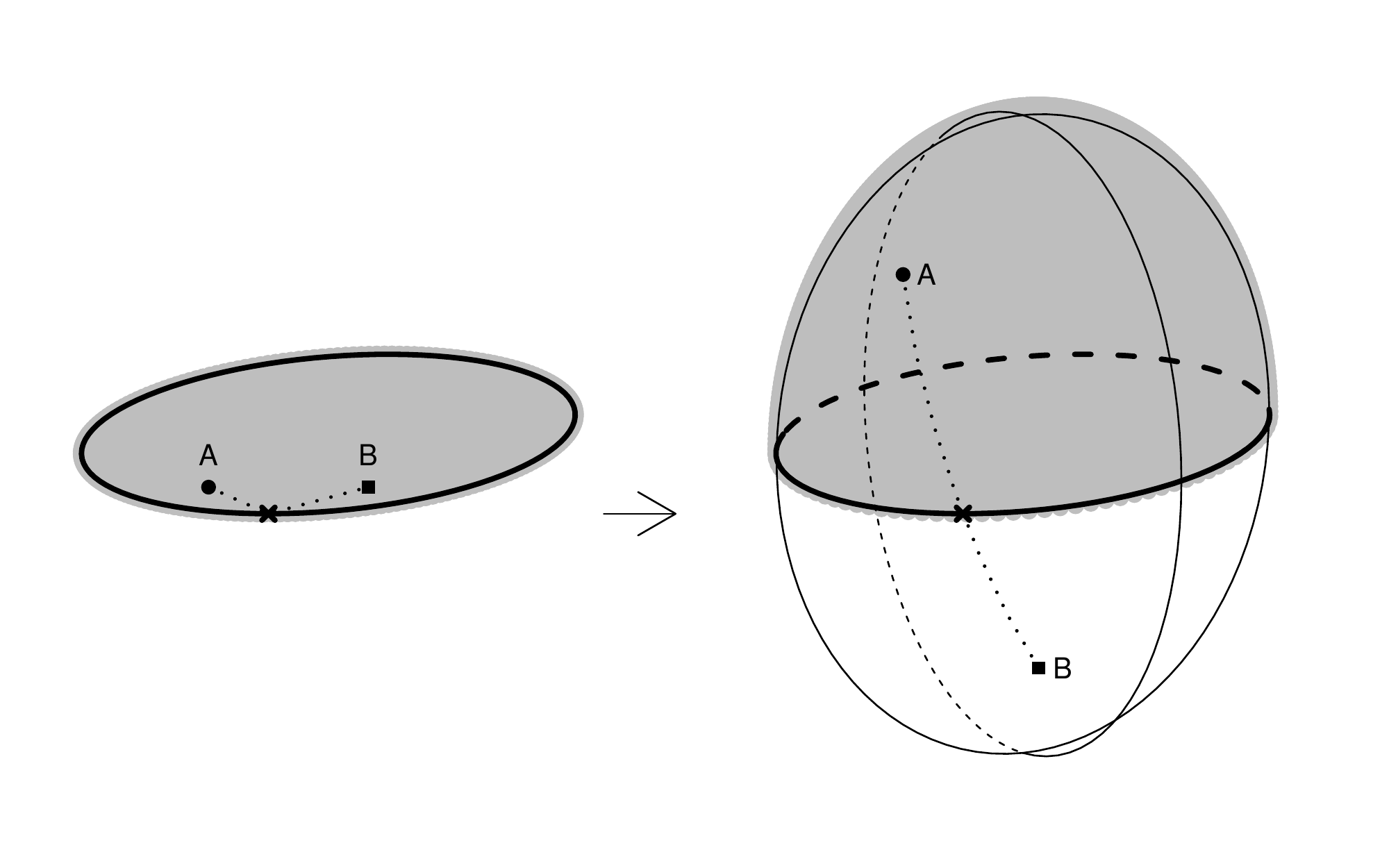}}
\caption{Transforming unit ball ${\bf B}_0^D(1)$ to sphere ${\bf S}^D$.}
\vspace{-15pt}
\label{fig:B2S}
\end{center}
\end{figure}

Using the above transformation, we define the dynamics on the sphere. This way, the resulting HMC sampler can move freely on ${\bf S}^D$ while implicitly handling the constraints imposed on the original parameters. As illustrated in Figure \ref{fig:B2S}, the boundary of the constraint, i.e., $\Vert\theta\Vert_2=1$, corresponds to the equator on the sphere ${\bf S}^D$. Therefore, as the sampler moves on the sphere, passing across the equator from one hemisphere to the other translates to ``bouncing back'' off the the boundary in the original parameter space. 

We have shown that by defining HMC on the sphere, besides handling the constraints implicitly, the computational efficiency of the sampling algorithm could be improved since the resulting dynamics has a partial analytical solution (geodesic flow on the sphere). We use this approach, called Spherical HMC, for sampling from the posterior distribution of $\beta$'s in the copula model. See Algorithm \ref{sphHMC} in Appendix for more details. 

Using parallelization (i.e., assigning each neuron to a worker to sample the parameters of the corresponding Gaussian process model), our computational method can handle relatively large numbers of neurons. For 10 neurons, 20 trials, and 50 time bins, each iteration of MCMC takes 8.4 seconds with acceptance probability of 0.72. For 50 neurons, the time per iteration increases to 24.5 with similar acceptance probability. 

All computer programs and simulated datasets discussed in this paper are available online at \url{http://www.ics.uci.edu/~babaks/Site/Codes.html}.

\section{Discussion} \label{discussion}
The method we proposed in this paper benefits from multiple aspects, including flexibility, computational efficiency, interpretability, and generalizability. The latter is especially important because the model offered in this work can be adopted to other computationally intensive biological problems.

The sampling algorithm proposed for detecting synchrony among multiple neurons is advantageous over commonly used MCMC techniques such as Metropolis-Hastings. This fact becomes even more salient given that current technology provides high-dimensional data by allowing the simultaneous recording of many neurons. Developing efficient sampling algorithms for such problems has also been discussed by \cite{ahmadian11}.

The analysis results presented here have demonstrated a number of ways in which examining the temporal relationship of activity in multiple neurons can reveal information about population dynamics of neuronal circuits. These kinds of data are critical in going beyond treating populations as averages of single neurons, as is commonly done in physiological studies. They also allow us to ask the question of whether neuronal firing heterogeneity contributes towards a unified whole \citep{stokes13} or whether separate populations in a brain area are differentially involved in different aspects of behavior \citep{buschman12}.

In our current model, $\beta$ are fixed. However, dependencies among neurons could in fact change over time. To address this issue, we could allow $\beta$'s to be piecewise constant over time to capture non-stationary neural connections. Change-point detection would of course remain a challenge.

\clearpage

\section*{Appendix}

\appendix

\begin{algorithm}[h]
\begin{algorithmic}
\caption{Sampling latent variables, copula parameters, and hyperparameters}
\label{mainAlg}

\STATE Initialize the matrix of latent variables, $U |_{(n,D)}$, where the $i^{th}$ column corresponds to the latent variables of the $i^{th}$ neuron, $n$ is the number of time bins, and $D$ is the number of neurons.
\STATE Initialize the hyperparameters, $\theta$, which specify the Gaussian process priors for the latent variables.
\STATE Initialize the copula model parameters, $\beta$, as a $D(D-1)/2$ vector.
\FOR{$i = 1, \ldots, B$}
    \STATE Sample $U^{(i+1)}$ from posterior distribution conditional on $U^{(i)}$, $\theta^{(i)}$ and $\beta^{(i)}$, $P(U^{(i+1)}|Y,U^{(i)},\theta^{(i)},\beta^{(i)})$, using the elliptical slice sampler (Algorithms \ref{ellipSlice}).\\
    \FOR{$j = 1, \ldots, D$}
        \STATE Sample $\theta_{j}^{(i+1)}$ from the posterior distribution of the hyperparameters of the $j^{th}$ latent variable conditional on the latent variables, $P(\theta_{j}^{(i+1)}|U_{j}^{(i+1)}, \theta_{j}^{(i)})$, using the slice sampler \citep{neal03}.\\
    \ENDFOR
    \STATE Sample $\beta^{(i+1)}$ from the posterior distribution conditional on the latent variables, $P(\beta^{(i+1)}|Y, U^{(i+1)}, \beta^{(i)})$, using Spherical HMC presented (Algorithm \ref{sphHMC}).
\ENDFOR
\end{algorithmic}
\end{algorithm}

\begin{algorithm}[h]
\begin{algorithmic}
\caption{Elliptical slice sampler for latent variables}\label{ellipSlice}
\STATE Let $U$ be the current state of the latent variables.
\STATE Sample $U^* \sim N(0,\Sigma)$, where $\Sigma$ is the covariance matrix of the Gaussian process.
\STATE Calculate the log-likelihood threshold for the elliptical slice sampler,
\STATE $v \sim \text{Uniform}[0,1] $
\STATE $ \log y \leftarrow \log(L(U)) + \log(v)$
\STATE Let $\alpha$ be the angle for the slice. 
\STATE Draw a proposal and define the corresponding bracket,
\STATE $\alpha \sim \text{Uniform}[0,2\pi]$
\STATE $(\alpha_{\min},\alpha_{\max}) \leftarrow (\alpha-2\pi,\alpha)$
\STATE Set $U' \leftarrow U \cos(\alpha) + U^*\sin(\alpha)$
\WHILE{$ \log(L(U')) < \log y $}
\IF{$\alpha < 0$} 
\STATE $\alpha_{\min} \leftarrow \alpha$
\ELSE 
\STATE $\alpha_{\max} \leftarrow \alpha$
\ENDIF
\STATE $\alpha \sim \text{Uniform}[\alpha_{\min},\alpha_{\max}]$
\STATE $U' \leftarrow U \cos(\alpha) + U^*\sin(\alpha)$
\ENDWHILE
\STATE Return $U'$ as the new state.
\end{algorithmic}
\end{algorithm}

\clearpage

\begin{algorithm}[h]
\caption{Spherical HMC for copula parameters}
\label{sphHMC}
\begin{algorithmic}
\STATE Initialize the copula parameters, $\beta^{(1)}$, along with their appropriate transformation, $\tilde\beta^{(1)}$, at the current state.
\STATE Sample a new momentum value $\tilde v^{(1)}\sim \mathcal N(0,I_{D+1})$.
\STATE Define the potential energy, $U$, as minus log density of $\tilde \beta$ and the kinetic energy, $K$, as minus log density of $\tilde v$.
\STATE Set $\tilde v^{(1)} \leftarrow \tilde v^{(1)} - \tilde\beta^{(1)} (\tilde\beta^{(1)})^T \tilde v^{(1)}$
\STATE Calculate the Hamiltonian function: $H(\tilde\beta^{(1)},\tilde v^{(1)})=U(\tilde \beta^{(1)}) + K(\tilde v^{(1)})$ 
\FOR{$\ell=1$ to $L$}
\STATE $\tilde v^{(\ell+\frac{1}{2})} = \tilde v^{(\ell)}-\frac{\epsilon}{2} \left(\begin{bmatrix} I_D\\ 0\end{bmatrix} -\tilde\beta^{(\ell)} (\beta^{(\ell)})^T\right) \nabla U(\beta^{(\ell)})$
\STATE $\tilde \beta^{(\ell+1)} = \tilde \beta^{(\ell)} \cos(\Vert \tilde v^{(\ell+\frac{1}{2})}\Vert \epsilon) + \frac{\tilde v^{(\ell+\frac{1}{2})}}{\Vert \tilde v^{(\ell+\frac{1}{2})}\Vert} \sin(\Vert \tilde v^{(\ell+\frac{1}{2})}\Vert \epsilon)$
\STATE $\tilde v^{(\ell+\frac{1}{2})} \leftarrow -\tilde\beta^{(\ell)}\Vert \tilde v^{(\ell+\frac{1}{2})}\Vert \sin(\Vert \tilde v^{(\ell+\frac{1}{2})}\Vert \epsilon)$
\STATE $\phantom{\tilde v^{(\ell+\frac{1}{2})} \leftarrow} + \tilde v^{(\ell+\frac{1}{2})} \cos(\Vert \tilde v^{(\ell+\frac{1}{2})}\Vert \epsilon)$
\STATE $\tilde v^{(\ell+1)} = \tilde v^{(\ell+\frac{1}{2})}-\frac{\epsilon}{2} \left(\begin{bmatrix} I_D\\ 0\end{bmatrix} -\tilde\beta^{(\ell+1)} (\beta^{(\ell+1)})^T\right) \nabla U(\beta^{(\ell+1)})$
\ENDFOR
\STATE Calculate $H(\tilde\beta^{(L+1)},\tilde v^{(L+1)})=U(\tilde \beta^{(L+1)}) + K(\tilde v^{(L+1)})$
\STATE Calculate the acceptance probability $$\alpha = \exp\{-H(\tilde\beta^{(L+1)},\tilde v^{(L+1)})+H(\tilde\beta^{(1)},\tilde v^{(1)})\}$$
\STATE Accept or reject the proposal according to $\alpha$
%\STATE Calculate the corresponding weight $|\beta_{D+1}^{(n)}|$
\end{algorithmic}
\end{algorithm}

\clearpage

\bibliographystyle{apa}

\end{document}